\documentclass[aps,pre,floatfix,nofootinbib,showpacs]{revtex4} 
\usepackage{graphicx} \usepackage{amsmath} \usepackage{amssymb}
\newcommand{\BEQ}{\begin{equation}}
\newcommand{\EEQ}{\end{equation}}
\newcommand{\BEA}{\begin{eqnarray}}
\newcommand{\EEA}{\end{eqnarray}}
\newcommand{\BGA}{\begin{gather}}
\newcommand{\EGA}{\end{gather}}

\renewcommand{\d}{{\rm d}}
\newcommand{\eps}{\eta}
\newcommand{\epf}{\varphi}

\newcommand{\spe}{l}
\newcommand{\p}{q }
\newcommand{\al}{\alpha }

\newcommand{\e}{\epsilon }

\newcommand{\x}{x }
\newcommand{\bx}{{\bf x}_{N\ldots 1} }
\newcommand{\bfx}{{\bf x} }
\newcommand{\bix}{{\bf x}_{1\ldots} }
\newcommand{\X}{{\cal X} }

\renewcommand{\S}{{\cal S} }

\newcommand{\T}{\mathbb{T}}
\renewcommand{\P}{\mathbb{P}}
\renewcommand{\l}{\lambda^n}

\begin{document}
\draft
\title
{Entropy of Hidden Markov Processes via Cycle Expansion.}

\date{\today}
\author{Armen E. Allahverdyan}
\affiliation{Yerevan Physics Institute,
Alikhanian Brothers Street 2, Yerevan 375036, Armenia}

\begin{abstract} Hidden Markov Processes (HMP) is one of the basic tools
of the modern probabilistic modeling. The characterization of their
entropy remains however an open problem. Here the entropy of HMP is
calculated via the cycle expansion of the zeta-function, a method
adopted from the theory of dynamical systems. For a class of HMP this
method produces exact results both for the entropy and the
moment-generating function.  The latter allows to estimate, via the
Chernoff bound, the probabilities of large deviations for the HMP. More
generally, the method offers a representation of the moment-generating
function and of the entropy via convergent series.

\end{abstract}


%%Information theory, 89.70.-a
%%channel capacity in, 89.70.Kn
%%entropy in, 89.70.Cf
%%05.20.-y Classical statistical mechanics 

\maketitle

\section{Introduction.}

Hidden Markov Processes (HMP) are generated by a Markov process observed
via a memory-less noisy channel. They are widely employed in various
areas of probabilistic modeling
\cite{rabiner_review,ephraim_review,signal,dna}: information theory,
signal processing, bioinformatics, mathematical economics, linguistics,
{\it etc}.  One of the main reasons for these numerous applications is
that HMP present simple and flexible models for a history-dependent
random process. This is in contrast to the Markov process, where the
history is irrelevant, since the future of the process depends on its
present state only. 

Much attention was devoted to the entropy of HMP
\cite{ash,cover_thomas,strat,blackwell,reza,birch,jacquet,gold,weissman,egner,zuk_jsp,zuk_aizenman,chigan}.
It characterizes the information content (minimal number of bits needed
for a reliable encoding) of HMP viewed as a probabilistic source of
information. More specifically, the realizations generated in the long
run of a random ergodic process, e.g. HMP, are divided into two sets
\cite{cover_thomas,strat}. The first (typical) set is the 
smallest set of realizations with the overall probability close to
one. The rest of realizations are contained in the second,
low-probability set. Now the entropy characterizes the number of
elements in the typical set \cite{cover_thomas,strat}. When HMP is
employed as a model for information transmission over a noisy channel,
the entropy is still important, since it is the basic non-trivial
component of the channel capacity (other components needed for
reconstructing the channel capacity are normally easier to calculate and
characterize). 

However, there is no direct formula for the entropy of HMP, in contrast
to the Markov case where such a formula is well-known
\cite{ash,cover_thomas,strat}. Thus people studied the entropy via
expansions around various limiting cases, or via upper and lower bounds
\cite{cover_thomas,birch,jacquet,gold,weissman,egner,zuk_jsp,zuk_aizenman,chigan}.
There is also a general formalism that expresses the entropy of HMP via
the solution of an integral equation \cite{strat,blackwell,reza}. This
formalism is however relatively difficult to apply in practice. 

Once the entropy characterizes the number of typical long-run
realizations, it is of interest to estimate the probability of atypical
realizations. These estimates are standardly given via the
moment-generating function of the random process
\cite{cover_thomas,strat}.  The knowledge of this function also allows
to reconstruct the entropy \cite{cover_thomas,strat}.

This paper presents a method for calculating the moment-generating
function of HMP. The method is adopted from the theory of chaotic
dynamical systems, where it is known as the cycle expansion of the
zeta-function \cite{artuso,mainieri}. We show that in a certain class of
HMP one can obtain exact expressions for the moment-generating function
and for the entropy. For other cases the method offers analytic
approximations of the moment-generating function via 
convergent power series. 

We attempted to make this paper self-contained and organized it as
follows. Section \ref{def} defines the HMP, settles some notations, and
recalls how to express the probabilities of HMP via a random matrix
product. In section \ref{ergo_entropy} we briefly review the main facts
about the entropy of an ergodic process and the corresponding typical
(highly probable) set of realizations. The main purpose of section
\ref{oseledec} is to relate the entropy of HMP to the spectral radius of
the corresponding random matrix product. This is done via the Lyapunov
exponent of the random matrix product.  Section \ref{daos} discusses the
moment-generating function of HMP.  This function is employed (via Chernoff bounds) for
characterizing the atypical (improbable) realizations of HMP.  Section
\ref{zeta_section} shows how to calculate the entropy and the generating
function via the zeta-function and the periodic orbit expansion.
Section \ref{aggr} discusses one of the simplest examples of HMP and
presents exact expressions for its entropy and the moment-generating
function. Here we also apply the moment-generating function for
estimating atypical realizations of the HMP. Section \ref{bshmp}
studies another popular model for HMP, binary symmetric HMP. It is
shown that the presented approach reproduces known approximate results
and predicts several new ones.  The last section shortly summarizes the
obtained results. Some issues, which are either too technical or too
general for the present purposes, are discussed in Appendices. 

\section{Definition of the Hidden Markov Process.}
\label{def}

In this section we recall the definition of the Hidden Markov Process
(HMP); see \cite{rabiner_review,ephraim_review} for reviews. 

Let a discrete-time random process $\S=\{\S_0,\S_1,\S_2,...\}$  
be Markovian, with time-independent conditional probability 
\BEA
{\rm
Pr}[\S_{k}=s_{k}|\S_{k-1}=s_{k-1}] ={\rm Pr}[\S_{k+l}=s_{k}|\S_{k-1+l}=s_{k-1}]
=p(s_{k}|s_{k-1}),
\EEA
where $l$ is an integer. Each realization $s$ of the
random variable $\S$ takes values $s=1,...,L$. The joint 
probability of the Markov process reads
\BEA
\label{0}
{\rm Pr}[\S_N=s_N,...,\S_0=s_0]=p(s_{N}|s_{N-1})\ldots p(s_{1}|s_{0})p(s_{0})=
\prod_{k=N}^1 p(s_{k}|s_{k-1})\,
p(s_{0}),
\EEA
where $p(s_0)$ is the initial probability.
The conditional probabilities $p(s_{k}|s_{k-1})$ 
define the $L\times L$ transition matrix $\P$:
\BEA
\label{makarov}
\P_{s_{k}\,s_{k-1}}= p(s_{k}|s_{k-1}).
\label{trans_mat}
\EEA

We assume that the Markov process $\S$ is mixing \cite{horn}:
it has a unique stationary distribution $p_{\rm st}(s)$,
\BEA
\label{graal}
\sum_{s'=1}^L p(s|s') p_{\rm st }(s')=p_{\rm st }(s),
\EEA
that is established from any initial probability in the long time limit.
The transition matrix $\P$ has always one eigenvalue equal to $1$ [since $\P$
has a left eigenvector $(1,...,1)$], and the modules [absolute
values] of all other eigenvalues are not larger than one
\footnote{Indeed, $\sum_k \P_{ik}x_k=\nu x_i$ implies $|\sum_k
\P_{ik}x_k|\leq \sum_k \P_{ik}|x_k| =|\nu| |x_i|$, which then leads
to $|\nu|\leq 1$.}. The mixing feature however demands that the
eigenvalue equal to $1$ is non-degenerate and the modules of all other
eigenvalues are smaller than $1$ \cite{horn}.  A sufficient condition
for mixing is that all the conditional probabilities $p(s_{i+1}|s_{i})$
of the Markov process are positive \cite{horn} \footnote{
Weaker sufficient conditions for mixing are that {\it i)}
for any $(i,j)$ there exists a positive integer
$m_{ij}$ such that $(\P^{m_{ij}})_{ij}>0$, i.e., for
some power of the matrix its entries are positive, and {\it ii)} $\P$
has at least one positive diagonal element \cite{horn}. If we do assume the first condition, but
do not assume the second one, the eigenvalue $1$ of $\P$ is [algebraically and thus geometrically]
non-degenerate, and is not smaller than the absolute values of all other
eigenvalues \cite{horn}. The corresponding [unique] eigenvector has strictly positive
components. However, it may be that the module of some other
eigenvalue(s) is equal to $1$ thus preventing
the proper mixing, but still allowing for ergodicity due to condition {\it i)}.
}.  Taking $p(s)=p_{\rm st}(s)$ in
(\ref{0}) makes the process $\S$ stationary. 

Let random variables $\X_i$, with realizations $\x_i=1,..,M$, be noisy
observations of $\S_i$: the (time-invariant) conditional probability of
observing $\X_i=\x_i$ given the realization $\S_i=s_i$ of the Markov
process is $\pi(\x_k|s_k)$. The joint probability of the original process 
and its noisy observations reads
\BEA
\label{1}
P(s_{N},\ldots,s_0;\x_N,\dots,\x_1)
&=& \prod_{k=N}^1 \pi(\x_k|s_k)p(s_{k}|s_{k-1})
p_{\rm st}(s_{0})\\
&=& T_{s_{N}\,s_{N-1}}(\x_{N})...
T_{s_{1}\,s_{0}}(\x_1)\, p_{\rm st}(s_0), 
\EEA
where the $L\times L$ transfer-matrix $T(x)$ with matrix elements 
$T_{s_{i}\,s_{i-1}}(\x)$ is defined as
\BEA
\label{transfer}
T_{s_{i}\,s_{i-1}}(\x)= \pi(\x|s_i)\, p(s_{i}|s_{i-1}).
\EEA

Thus $\X=\{\X_1,\X_2,...\}$, called hidden Markov process, results from
observing the Markov process $\S$ through a memory-less process with the
conditional probability $\pi(\x|s)$.  The composite process $\S\X$ is
Markovian as well. 

The probabilities for the process $\X$ are represented via 
the transfer matrix product (similar representation were employed in
\cite{jacquet,gold})
\BEA
\label{mubarak}
\label{2}
&&P(\bx)
=\langle {\rm un}|\T(\bx) |{\rm st}\rangle, \\
\label{barak}
&&\T(\bx)\equiv\prod_{k=N}^1 T(\x_k), \\ 
&&\bx \equiv (\x_N,...,\x_1),
\label{laokon}
\EEA
where we used the bra(c)ket notations: $|{\rm st}\rangle $ is the column
vector with elements $p_{\rm st}(k)$, $k=1,...,L$, and  $\langle
{\rm un}|=(1,...,1)$. 

The HMP defined by (\ref{2}) is (in general) not a Markov
process, i.e., its probabilities do not factorize as in (\ref{0}). Thus
the history of the process can become relevant. This is the underlying
reason for widespread applications of HMP. 

The process $\X$ is stationary due to the stationarity of $\S$:
\BEA
\label{sato}
{\rm Pr}[\X_{N+l}=\x_{N},...,\X_{l+1}=\x_{1}]
={\rm Pr}[\X_{N}=\x_{N},...,\X_{1}=\x_{1}]
=P(\x_N,...,\x_1),
\EEA
where $l$ is a positive integer. 

In addition, $\X$ inherits the mixing feature from the underlying Markov process $\S$
\cite{ephraim_review}, because the observation process by itself is
memoryless: $\pi(\x_k|s_k)= \pi(\x_k|s_k,s_{k-1}, ..., s_0)$. 
(The general definitions of ergodicity and mixing are reminded below.)

\subsection{Notations for the eigenvalues and singular values.}
\label{nota}

For future purposes we concretise some notations.  For a matrix $A$, let
$\spe_0[A], \spe_1[A], ....$ be the modules of its eigenvalues.
We order $\spe_k[A]$ as
\BEA
\label{karamba}
\lambda[A]\equiv\spe_0[A]
\,\geq\, \spe_1[A]\,\geq \ldots ,
\label{karamba_0}
\EEA
$\lambda[A]$ is called the spectral radius of 
$A$ \cite{horn}. If $A$ has non-negative matrix
elements, the spectral radius is an eigenvalue by itself \cite{horn}. 
Here are two obvious features of the function $\lambda$ ($d$ is a positive integer):
\BEA
\label{croat1}
&&\lambda[A^d]=(\lambda[A])^d, 
\\ &&\lambda[AB]=\lambda[BA],
\label{croat2}
\EEA
where (\ref{croat2}) follows from the fact that $AB$ and $BA$ have 
identical eigenvalues: $AB|\psi\rangle =\nu
|\psi\rangle$ implies $BA\, \left(B|\psi\rangle\right) =\nu
B|\psi\rangle$. 

Let $A^\dagger$ be the complex conjugate of $A$.  The singular values
$\sigma_k[A]\geq 0$ for a matrix $A$ are the eigenvalues of a hermitean
matrix $\sqrt{AA^\dagger}$ or, equivalently, of $\sqrt{A^\dagger A}$;
see Appendix \ref{app_Weyl} for a brief reminder on the features of the
singular values.  We order $\sigma_k[A]$ as
\BEA
\sigma_0[A]\geq \sigma_1[A]\geq....
\label{grad}
\EEA

\section{Entropy and typical set of ergodic processes.}
\label{ergo_entropy}

The $N$-block entropy of a stationary [not necessarily Hidden Markov] 
random process $\X$ is defined as \cite{ash,cover_thomas,strat}
\BEA
\label{block}
H(N)=H(\X_1,...,\X_N)\equiv -\sum_{\bx} P(\bx) \ln P(\bx),
\EEA
where the probability $P(\bx)$ is given as in (\ref{mubarak}), and where
$\bx$ is defined in (\ref{laokon}).
Various features of $H(N)$ and of several related quantities 
are discussed in Appendix \ref{app_entropy}.

Using (\ref{block}) one now defines the entropy (rate) of the random process
$\X$ as \cite{ash,cover_thomas,strat}
\BEA
h={\rm lim}_{N\to\infty}\frac{H(N)}{N}.
\label{hopo}
\EEA
Alternative representations of $h$ are recalled in Appendix \ref{app_entropy}.
In particular, $h$ is the uncertainty [per unit of time] of the random process given its long
history.

For ergodic processes the above definition of entropy can be related to a single, long
sequence of realizations \cite{ash,cover_thomas,strat}. First of all let us recall that
the process $\X$ is ergodic if it satisfies to the weak law of large numbers (time average
is equal to the space average): for any 
function $f$ with a finite expectation value $\bar{f}\equiv 
\sum_{\x_k,...,\x_0}f[\x_k,...,\x_0] P(\x_k,...,\x_0)$, we have 
probability-one convergence for $N\to \infty$
\cite{ash,cover_thomas,strat}:
\BEA
\label{blef2}
\frac{1}{N}\sum_{n=0}^{N-1}
f[\X_{n+k},...,\X_{n}]
\rightarrow\bar{f},
\EEA
i.e., for any positive numbers $\varepsilon$ and $\delta$,
there is such an integer ${\cal N}(\varepsilon, \delta)$ 
that for all $N>{\cal N}(\varepsilon, \delta)$,
\BEA
{\rm Pr}\left[
\left|
\frac{1}{N}\sum_{n=0}^{N-1}
f[\X_{n+k},...,\X_{n}] - \bar{f}
\right|\geq \varepsilon
\right]\leq \delta.
\EEA
Several alternative definitions of ergodicity are discussed
in \cite{karl}\footnote{\label{para}
One such definition is worth mentioning: $\X$ is ergodic if for any $k$, $m$
and $s$:
${\rm lim}_{N\to\infty}
\frac{1}{N}\sum_{n=0}^{N-1}
{\rm Pr}[\X_{n+k}=\x_{k},...,\X_{n}=\x_{0},\X_{m+s}=y_{m},...,\X_{s}=y_{0} ]
=P(\x_k,...,\x_0)P(y_m,...,y_0)$.
This definition admits a
straightforward and important generalization. $\X$ is called mixing if
the above relation holds without the time-averaging $\frac{1}{N}\sum_{n=0}^{N-1}$, but 
in the limit $n\to\infty$.
}.

Now the McMillan lemma states that for an ergodic process the entropy (\ref{hopo}) characterizes
individual realizations in the sense of probability-one convergence for $N\to\infty$
\cite{ash,cover_thomas,strat}
\footnote{The McMillan lemma contains two essential steps \cite{ash}. First is to realize that although
the definition (\ref{blef2}) of ergodicity does not apply directly to $\frac{1}{N}\ln P(\bx)$, it
does apply to the probability $Q_m(\bx)=P(x_1,...,x_m)\prod_{i=1}^{N-m}P(x_{m+i}|x_{m+i-1},...,x_i)$,
which defines an approximation of the original ergodic process by a $m$-order Markov process.
In the second step
using a chain of inequalities ${\rm Pr}[|\ln x|\geq n\varepsilon]
\leq \frac{1}{n\varepsilon}\overline{|\ln x|}\leq \frac{1}{n\varepsilon}\overline{(2x -\ln x)}$,
one proves that for any stationary [not necessarily ergodic] process
$Q_m(\bx)$ is indeed a good approximation in the sense of $\frac{1}{N}\ln\frac{Q_m(\bx)}{P(\bx)}\simeq 0$
for $N\gg m\to \infty$.}:
\BEA
-\frac{1}{N}\ln P(\bx)\rightarrow h \quad {\rm or} \quad 
{\rm Pr}\left[
\left| - \frac{1}{N}\ln P(\bx)- h
\right|\leq \varepsilon
\right]\geq 1- \delta. 
\label{desp}
\EEA
Based on (\ref{desp}) one defines the typical set
$\Omega^*_N(\varepsilon)$ as the set of all $\bx$, which satisfy to
\BEA
\label{typo}
h-\varepsilon \leq -\frac{1}{N}\ln P(\bx) \leq h+\varepsilon.
\EEA
Now (\ref{desp}) implies that ${\rm Pr}[\bx \in \Omega^*_N(\varepsilon)
]\geq 1-\delta$, i.e., the overall probability of $\Omega^*_N(\varepsilon)$ converges to one
in the limit $N\to\infty$. Since all elements in $\Omega^*_N(\varepsilon)$ have approximately
equal probabilities, the number of elements 
$|\Omega^*_N(\varepsilon)|$
in $\Omega^*_N(\varepsilon)$ scales as $e^{N h}$. More precisely,
this number  is estimated from (\ref{desp}, \ref{typo}) as \cite{ash}
\BEA
\label{bar_i}
(1-\delta)e^{N(h-\varepsilon)}\leq |\Omega^*_N(\varepsilon)|\leq e^{N(h+\varepsilon)}.
\EEA
Relations similar to (\ref{typo}) will be frequently
written as 
\BEA
P(\bx)\simeq e^{-Nh}\quad {\rm for} \quad \bx\in\Omega_N^*, 
\label{asym}
\EEA
meaning that the precise sense of the
asymptotic relation $\simeq$ for $N\to\infty$ can be clarified upon
introducing proper $\epsilon$ and $\delta$.

\section{Lyapunov exponents and entropy.}
\label{oseledec}

The purpose of this section is to establish relation (\ref{bobo_1})
between the entropy of a Hidden Markov Process, and the spectral radius
of the associated random matrix product (\ref{mubarak}). The reader may
skip this section, if this relation is taken granted. 

\subsection{Singular values of the random-matrix product.}
\label{yahud}

The actual calculation of the entropy $h$ for non-Markov processes meets
(in general) considerable difficulties. (For Markov processes definition
(\ref{hopo}) applies directly leading to the well-known formula for the
entropy \cite{ash}.) The first step in calculating the entropy $h$ for a
Hidden Markov Process (HMP) is to relate $h$ to the
large-$N$ behaviour of the $L\times L$ matrix $\T(\bx)$, 
which defines the probability of HMP; see (\ref{mubarak}, \ref{barak}).  Recall that
$\T(\bx)$ is a function of the random process $\X$. Assume that 
{\it i)} $\X$ is stationary, as is the case after (\ref{sato}).
{\it ii)} The average logarithm of the maximal singular value of $T(x)$
is finite: $\langle\ln \sigma_0 [T(x)] \rangle<\infty$. {\it iii)} $\X$
is ergodic. Then the subadditive ergodic theorem applies 
claiming for $N\to\infty$ the probability-one convergence \cite{kingman,steele}:
\BEA
\label{puk}
-\frac{1}{N}\ln\sigma_k[\mathbb{T}(\bx)]\rightarrow \mu_k,\quad k=0,\ldots, L-1,
\EEA 
where $\sigma_k[\mathbb{T}(\bx)]$ are the singular values of $\mathbb{T}(\bx)$ (see 
section \ref{nota} for notations),
and where $\mu_k$ are called Lyapunov exponents. According to (\ref{grad}) they 
are ordered as $\mu_0\leq \mu_1\leq ...$. 

Using the definition (\ref{typo}) of the typical set, (\ref{puk}) can be
written as an asymptotic relation
$\sigma_k[\mathbb{T}(\bx)]\simeq e^{-N\mu_k}$ for $\bx\in \Omega_N$ and
sufficiently large $N$ \cite{crisanti}.  Moreover, 
employing the singular value decomposition [see
Appendix \ref{app_Weyl}], one represents $\T(\bx)$ for $N\to\infty$ and
$\bx\in\Omega_N^*$ as
\BEA
\label{lelak}
\T(\bx)\simeq {\rm diag}\left[e^{-N \mu_0}, \ldots, e^{-N \mu_{L-1}}
\right]\, U({\bf x}), 
\EEA
where ${\rm diag}\left[a, \ldots, b\right]$ is a diagonal matrix with
entries $a,\ldots,b$, and where $U({\bf x})$ is an orthogonal matrix. 
The fact that (for $N\to\infty$) the matrix $U$ does not depend on $N$
(but does in general depend on the realization $\bfx$) is a consequence of the
Oseledec theorem \cite{crisanti,goldsheid}.

Thus the meaning of (\ref{lelak}) is that the essential dependence of
$\T(\bx)$ on $N$ is contained in the singular values $e^{-N
\mu_k}$, while $U(\bfx)$ does not depend on $N$ for $N\to\infty$. 

\subsection{Eigenvalues of the random-matrix product.}
\label{golem}

The above reasoning by itself is silent about the eigenvalues of
$\T(\bx)$. Since the matrix $\T(\bx)$ is in general not normal, i.e.,
the commutator of $\T(\bx)$ with its transpose $\T^\dagger(\bx)$ is not
zero, the modules $\spe_k[\T(\bx)]$ of its eigenvalues are not
automatically equal to its singular values $e^{-N\mu_k}$; see Appendix
\ref{app_Weyl}. For us the knowledge of the spectral radius
$\lambda[\T(\bx)]$ will be important, because for calculating the
entropy we shall employ a method that essentially relies on the features
(\ref{croat1}, \ref{croat2}), which hold for the eigenvalues, but do not
hold for singular values. 

It is shown in Appendix \ref{kovtun} that the
representation (\ref{lelak}) can be used for deducing that in the limit
$N\to\infty$ and for $\bx \in \Omega_N^*$ the spectral radius 
$\lambda[\T(\bx)]$ of $\T(\bx)$ behaves as
[recall (\ref{karamba})]
\BEA
\label{kond}
\lambda[\T(\bx)]\simeq e^{-N \mu_0},
\EEA
where $\mu_0$ is the so called top Lyapunov exponent. 
Appendix \ref{kovtun} discusses under which {\it generic}
conditions (\ref{kond}) holds; see also \cite{or} in this
context.

Using (\ref{mubarak}) we have asymptotically
for $N\to \infty$ and $\bx\in\Omega_N^*$
\BEA
\label{kav}
&& \T(\bx) \simeq  e^{-N\mu_0}
|\,R({\bf x})\,\rangle
\langle\, L(\bfx)\,|+{\cal O}[e^{-N\nu_1(\bx) }]    , \\
\label{fufu}
&& P(\bx)\simeq e^{-N\mu_0 +{\cal O}(1)}+{\cal O}[e^{-N\nu_1(\bx)}],
\EEA
where we denoted $\spe_1[\T(\bx)]\equiv e^{-N\nu_1(\bx)}$ [see
(\ref{karamba})], and where $|\,R(\bfx)\,\rangle$ and
$|\,L(\bfx)\,\rangle$ are, respectively, the right and left eigenvectors
of $\T(\bx)$; see Appendix \ref{app_Weyl}. They do not depend on $N$
(for $N\to\infty$) for the same reason as $U$ in (\ref{lelak}) does not
depend on $N$.  In writing down (\ref{kav}) we assumed that the spectral
radius $\lambda[\T(\bx)]$ is not a degenerate eigenvalue of $\T(\bx)$,
or at least that its algebraic and geometric degeneracies coincide (see
Appendix \ref{app_Weyl}). In that latter case one can then use (\ref{kav}) with
straightforward modifications and obtain (\ref{fufu}). 

The term ${\cal O}[e^{-N\nu_1(\bx)}]$ in (\ref{kav}, \ref{fufu}) can be
neglected for $N\to \infty$ provided that $\mu_0
>\nu_1(\bx\in\Omega_N^*)$.  The multiplicative correction ${\cal O}(1)$
in (\ref{fufu}) comes from the eigenvectors in
(\ref{kav}). This correction can be neglected if $\mu_0$ stays finite
for $N\to\infty$. Below we assume that these two hypotheses hold.
This implies
from (\ref{typo}) a straightforward relation between the entropy $h$ and 
the spectral radius $\lambda[\T(\bx)]$ of $\T(\bx)$:
\BEA
\label{bobo_1}
h=\mu_0={\rm lim}_{N\to\infty} \{\,-\frac{1}{N}\ln\lambda[\T(\bx)] \,\}.
\EEA

The relation between the top Lyapunov exponent and the entropy is
known \cite{jacquet,gold}. The above discussion
emphasizes the role of the spectral radius in this relation
\cite{mainieri}. 

\section{Generating function and atypical realizations}
\label{daos}

While the entropy characterizes typical realizations of the process, it
is of interest (mainly for a finite number of realizations) to describe
atypical realizations, those which fall out of the typical set
$\Omega_N^*$. 

To this end let us introduce the generating function \cite{strat}
\BEA
\label{4}
\Lambda^N(n,N)
=\sum_{\bx} \lambda^n \left[
\T(\bx)
\right],
\EEA
where $n$ is a non-negative number. (Note that $\Lambda^N(n,N)$ means $\Lambda(n,N)$
in degree of $N$.)

The generating function $\Lambda^N(n,N)$ 
is an analog of the partition sum in statistical physics \cite{strat}\,
\footnote{$\Lambda(n,N)$ is sometimes called the generalized Lyapunov exponent. It is
closely related to the concept of multi-fractality \cite{crisanti}.}. Writing
\BEA
\Lambda^N(n,N)=
\sum_{\bx\in\Omega^*_N} \lambda^n \left[ \T(\bx) \right]+
\sum_{\bx\not\in\Omega^*_N} \lambda^n \left[ \T(\bx) \right],
\label{bokacho}
\EEA
one notes that in the limits $N\to\infty$ and $n\to 1$ the second
contribution in the RHS of (\ref{bokacho}) can be neglected due to 
definition (\ref{typo}, \ref{asym}) of the typicality, and then
$\Lambda^N(n,N)=\Lambda^N(n)=e^{-(n-1)Nh}$; see (\ref{kav}, \ref{fufu}).  
Here we already noted that $\Lambda(n,N)$ does not depend on $N$ for $N\to \infty$,
and denoted (in this limit) $\Lambda(n,N)=\Lambda(n)$.

Taking into account that $\Lambda(1)=1$, the entropy $h$ is calculated via 
derivative of the generating function:
\BEA
\label{ad_din}
h&=&-\frac{1}{N}\left.\frac{\partial \Lambda^N(n)}{\partial n} \right|_{n=1}=-
\left.\frac{\d\Lambda(n)}{\d n}\right|_{n=1}\equiv 
-\Lambda'(1)\\
 &=& -\sum_{\bx} \lambda \left[ \T(\bx)
\right] \ln \lambda \left[
\T(\bx)
\right].
\EEA

The generating function (\ref{4}) can be employed for estimating the
weight of atypical sequences. This estimate is known as the Chernoff
bound \cite{cover_thomas,strat}, and now we briefly recall its
derivation adopted to our situation. 

Consider the overall weight of atypical sequences, which have
probability lower than the typical-sequence probability $e^{-Nh}$; see
(\ref{typo}, \ref{asym}).  These atypical sequences are defined to
satisfy
\BEA
\label{kamsar}
-\ln\lambda \left[ \T(\bx) \right]>(1+\eps)Nh,
\EEA
where $\eta>0$ quantifies the deviation from the typical behavior. 
Let $\overline{\sum}_{\bx}$ be the sum over all those $\bx$ that satisfy
to (\ref{kamsar}). Define an auxiliary probability distribution
$\widetilde{P}(\bx|n)=\Lambda^{-N}(n,N)\,\lambda^n \left[ \T(\bx)
\right]$. The sought weight of the atypical sequences 
is expressed as ($\eps>0$ and $0<n<1$):
\BEA
\overline{\sum}_{\bx}\lambda \left[ \T(\bx) \right]&=&
\Lambda^{N}(n,N)\overline{\sum}_{\bx} \widetilde{P}(\bx|n)\,
e^{(1-n)\ln\lambda\left[ \T(\bx) \right]}\nonumber\\
&\leq&
e^{\,N[\,\ln\Lambda(n,N)+(n-1)(1+\eps) h\,]}\,\,
\overline{\sum}_{\bx} \widetilde{P}(\bx|n)
\leq e^{\,N[\,\ln\Lambda(n,N)+(n-1)(1+\eps) h\,]}.
\label{mako}
\EEA
Eq.~(\ref{mako}) leads to the following upper (Chernoff) bound for the weight of atypical sequences
with the probability lower than the $e^{-Nh}$:
\BEA
&&\sum_{-\ln\lambda \left[ \T(\bx) \right]>(1+\eps)Nh}\lambda \left[ \T(\bx) \right]
\leq e^{-\,N\, f(\eps)},\\
&&f(\eps)\equiv{\rm max}_{\,0<n<1\,}[\,\ln\frac{1}{\Lambda(n)}+(1-n)(1+\eps) h\,], \quad \eps>0.
\label{gusi_f}
\EEA

Analogously to (\ref{mako}) we get for the weight of the atypical sequences with the
probability higher than the $e^{-Nh}$ ($0<\eps<1$):
\BEA
&&\sum_{-\ln\lambda \left[ \T(\bx) \right]<(1-\eps)Nh}\lambda \left[ \T(\bx) \right]
\leq e^{-\,N\, g(\eps)},\\
&&g(\eps)\equiv{\rm max}_{\,n>1\,}[\,\ln\frac{1}{\Lambda(n)}+(1-n)(1-\eps) h\,], \quad \eps>0.
\label{gusi_g}
\EEA

The functions $f(\eta)$ and $g(\eta)$ in (\ref{gusi_f}) and
(\ref{gusi_g}), respectively, are called the rate functions
\cite{cover_thomas}. It is seen that $f(\eta)$ and $g(\eta)$ are the
Legendre transforms of $\ln\Lambda(n)$. The latter is a convex function
of $n$, $\frac{\d^2}{\d^2 n}\ln\Lambda(n)\geq 0$, as follow from its
definition (\ref{4}). Then $f(\eta)$ and $g(\eta)$ are convex as well
\cite{strat}. For example taking into account that $n$ and $\eta$ are
related via the extremum condition $\frac{\d}{\d n}\ln\Lambda(n)
=-(1+\eta)h$, we get 
$f''(\eta)=\left(\frac{\d n}{\d \eta}\right)^2
\left[\frac{\d^2}{\d n^2}\ln\Lambda(n)\right]_{n=n(\eta)}
\geq 0$.

While the above reasoning is based on the Chernoff bounds, there is
another (related, but more formal) approach to describing atypical
realization, which is known as the measure concentration theory. For a
recent application of this theory to HMP see \cite{kontorovich}. 

\section{Zeta function and its expansion over the periodic orbits (cycles).}
\label{zeta_section}

\subsection{Zeta function and entropy.}

In this section we show how to adopt the method proposed in
\cite{artuso,mainieri} for calculating the moment-generating function
$\Lambda(n)$ (and thus for calculating the entropy $h$ via
(\ref{ad_din})).  The method is based on the concepts of the
zeta-function and periodic orbits. 

Define the inverse zeta-function as \cite{strat,artuso,aurell,ruelle}
\BEA
\label{zeta_meta}
\xi(z,n)=\exp\left[-
\sum_{m=1}^\infty \frac{z^m}{m}\Lambda^m(n,m)
\right],
\EEA
where $\Lambda^m(n,m)\geq 0$ is given by (\ref{4}). 
The analogs of (\ref{zeta_meta}) are well-known in the theory of dynamic
systems; see \cite{ruelle} for a mathematical introduction, and
\cite{artuso,mainieri,aurell} for a physicist-oriented discussion. 

Since for a large $N$, $\Lambda^N(n,N)\to \Lambda^N(n)$, the zeta-function
$\xi(z,n)$ has a zero at $z=\frac{1}{ \Lambda(n) }$: 
\BEA
\xi(\frac{1}{\Lambda(n)},\, n)=0.
\label{khrych}
\EEA
Indeed for $z$ close (but smaller than) $\frac{1}{ \Lambda(n) }$, the series
$\sum_{m=1}^\infty \frac{z^m}{m}\Lambda^m(n,m)\to 
\sum_{m=1}^\infty \frac{[z\Lambda(n)]^m}{m}$ almost diverges and
one has $\xi(z)\to 1-z\Lambda(n)$.

Recalling that $\Lambda(1)=1$ and taking $n\to 1$ in
\BEA
0=\frac{\d}{\d n}\xi (\frac{1}{\Lambda(n)},n)=-\frac{\Lambda'(n)}{\Lambda^2(n)}\,\,
\frac{\partial}{\partial z}\xi (\frac{1}{\Lambda(n)},n)+
\frac{\partial}{\partial n}\xi (\frac{1}{\Lambda(n)},n),
\EEA
we get for the entropy from (\ref{ad_din})
\BEA
\label{mkno}
h=-\Lambda'(1)=-\frac{\frac{\partial}{\partial n}\xi(1,1)}{\frac{\partial}{\partial z}\xi(1,1)}.
\EEA

\subsection{Expansion over the periodic orbits.}

In Appendix \ref{fidelito} we describe following to \cite{artuso,ruelle,mainieri,aurell} that under conditions
(\ref{croat1}, \ref{croat2}) one can expand $\xi(z,n)$ over the periodic orbits:
\BEA
\label{castro}
&&\xi(z,n)=\prod_{p=1}^\infty \,
\prod_{\Gamma_p \in {\rm Per}(p) }\, 
\left[
1-z^p \lambda^n[T(\x_{\gamma_1})...T(\x_{\gamma_p})]\,
\right], \\ 
&&\Gamma_p\equiv (\gamma_1,...,\gamma_p), 
\label{gesh}
\EEA
where $\gamma_i=1,...,M$ are the indices referring to the realizations
of the random process $\X$. The set of periodic orbits ${\rm
Per}(p)$ contains sequences $\Gamma_p=(\gamma_1,...,\gamma_p)$ selected
according to the following two rules: {\it i)} $\Gamma_p$ turns to
itself after $p$ successive cyclic permutations of its elements, but it
does not turn to itself after any smaller (than $p$) number of
successive cyclic permutations; {\it ii)} if $\Gamma_p$ is in ${\rm
Per}(p)$, then ${\rm Per}(p)$ contains none of those $p-1$ sequences
obtained from $\Gamma_p$ under $p-1$ successive cyclic permutations. Concrete
examples of ${\rm Per}(p)$ for $M=2,3$ are given in Tables \ref{tab1} and
\ref{tab2}. 

It is more convenient to present (\ref{castro}) as an infinite sum \cite{artuso,mainieri,gaspar}
\BEA
\label{retmon}
\xi(z,n)=1-z\sum_{l=1}^M \l_l
+\sum_{k=2}^\infty \epf_k(n) z^k,
\EEA
where we defined
\BEA
\label{depr}
\l_{\alpha...\beta}\equiv\l [T(\x_{\alpha})...T(\x_{\beta})], \qquad
\l_{\alpha+\beta}\equiv\l [T(\x_{\alpha})]\l[T(\x_{\beta})], 
\EEA
and where $\epf_k(n)$ are calculated from (\ref{castro}, \ref{gesh}) and recipes
presented in Appendix \ref{zeta_app}.  These calculations become tedious for large
values of $k$ in $\epf_k(n)$. This is why in Appendix \ref{mato} it is shown how to generate
$\epf_k(n)$ via Mathematica 5. 

For two ($M=2$) realizations of the HMP we employ the notations
(\ref{depr}) and get for the first few terms of the product (\ref{castro})
[consult Table \ref{tab1} for understanding the origin of these terms]
\BEA
\label{koriz}
\xi(z,n)&=&
(1-z\lambda^n_{1})\,(1-z\lambda^n_{2})\,(1-z\lambda^n_{12})\,
(1-z\lambda^n_{122})\,(1-z\lambda^n_{112})\\
&&(1-z\lambda^n_{1222})\,(1-z\lambda^n_{1112})\,(1-z\lambda^n_{1122})
\prod_{p=5}^\infty \,
\prod_{\Gamma_p \in {\rm Per}(p) }\, 
\left(
1-z^p \lambda^n_{\gamma_1...\gamma_p}
\right).
\EEA

For the first six terms of the expansion (\ref{retmon}) we get
\BEA
\epf_2(n)&=&-\l_{12}+\l_{1+2},
\label{jo2}
\EEA
\BEA
\epf_3(n)&=&-\l_{221}+\l_{2+21}-\l_{112}+\l_{1+12},
\label{jo3}
\EEA
\BEA
\label{jo4.1}
\epf_4(n)&=&-\l_{1122}+\l_{2+211}-\l_{1222}+\l_{2+122}
-\l_{1112}+\l_{1+211}\\
&&-\l_{1+2+12}+\l_{1+122}
\label{jo4.2}
\EEA
\BEA
\label{jo5.1}
\epf_5(n)&=&-\l_{11222}+\l_{1+1222}-\l_{11122}+\l_{2+1112}\\
\label{jo5.2}
&&-\l_{11112}+\l_{1+1112} -\l_{12222}+\l_{2+1222} \\
\label{jo5.3}
&&-\l_{12121}+\l_{1+1122}-\l_{12122}+\l_{2+1122}\\
\label{jo5.4}
&&-\l_{1+2+122}+\l_{12+122}-\l_{1+2+112}+\l_{12+112},
\EEA
\BEA
\label{jo6.1}
\epf_6(n)&=&-\l_{111122}+\l_{1+11122}-\l_{112122}+\l_{1+12122}-\l_{111222}+\l_{1+11222}\\
\label{jo6.2}
&&-\l_{111212}+\l_{1+11212}-\l_{112222}+\l_{1+12222}-\l_{222121}+\l_{2+22121}\\
\label{jo6.3}
&&-\l_{12 2222}+\l_{2+12 222}-\l_{111 112}+\l_{1+11 112}-\l_{112212}+\l_{2+12121}\\
\label{jo6.4}
&&-\l_{1+12+122}+\l_{1+12122}-\l_{2+12+211}+\l_{12+1122}-\l_{1+12+211}+\l_{12+2111}\\
\label{jo6.5}
&&-\l_{2+12+122}+\l_{12+1 22 2}-\l_{1+2+1222}+\l_{2+11 222}
-\l_{1+2+2 111}+\l_{2+21 111}\\
&&-\l_{1+2+1122}+\l_{122+211}.
\label{jo6}
\EEA

In section \ref{exo} we study examples, where the expansion
(\ref{retmon}) can be summed exactly. In these examples the sum in
(\ref{retmon}) exponentially convergences for $|z|<\alpha^n$, where
$\alpha> 1$ is a parameter. As discussed in \cite{aurell},
the exponential convergence of $\xi(z)$ is expected to be a general
feature, and it is supported by rigorous results on the
structure of the zeta-function. 

\subsubsection{The structure of $\epf_k(n)$.}
\label{structure}

Note that $\epf_k$ consists of even number of terms. The terms are
grouped in pairs, e.g., $[-\l_{221}+\l_{2+21}]+[-\l_{112}+\l_{1+12}]$
for $\epf_3$, and analogously for other $\epf_k$'s. Each pair has the
form $-\l_{A}+\l_{B}$, where $A$ and $B$ have the same number of symbols
$1$ and the same number of symbols $2$. This feature ensures that when
the spectral radius of the product is equal to the product of the
spectral radii, all the terms $\epf_k$ will vanish. Ultimately, this is
the feature that enforces the convergence of (\ref{retmon})
\cite{artuso,aurell}. Once it converges, we can approximate $\xi(z,n)$ by a
polynomial of a finite order. 

The set of pairs for each $\epf_k$ can be divided further into several
groups. The first group is formed by (\ref{jo2}) and (\ref{jo3}) for
$\epf_2$ and $\epf_3$, respectively, by (\ref{jo4.1}) for $\epf_4$, by
(\ref{jo5.1}--\ref{jo5.3}) for $\epf_5$, and by (\ref{jo6.1}--\ref{jo6.3})
for $\epf_6$. The pairs in this group have the form $-\l_{Al}+\l_{A+l}$,
where $l=1$ or $l=2$. If $A$ contains $m$ indices and if $m$ is large,
we expect $\ln\l_{A}={\cal O}(m)$ according to the 
discussion in section \ref{golem}. Then 
\BEA
\label{israel}
-\l_{Al}+\l_{A+l}\to 0 \qquad {\rm for } \qquad m\to\infty. 
\EEA

The second group is given by (\ref{jo4.2}) for $\epf_4$, (\ref{jo5.4})
for $\epf_5$, and by (\ref{jo6.4}, \ref{jo6.5}) for $\epf_6$. In this
second group the terms have the form
$-\l_{A+B+C}+\l_{A+BC}=\l_{A}(\l_{B+C}-\l_{BC})$.  Here the term
$(\l_{B+C}-\l_{BC})$ has the structure of the first group. 
For $B$ or/and $C$ containing a large number of indices, 
$(\l_{B+C}-\l_{BC})$ will go to zero. 

Finally the third group appears only for $k\geq 6$. For $k=6$ this group
has only one pair given by (\ref{jo6}).  The members of this third group
are of the form $-\l_{A+B+CD}+\l_{ABD+C}$. 

Let us return to (\ref{israel}), which holds, in particular, for $A$
consisting of the same type of indices (e.g., $A$ containing only
$1$'s).  Recalling our discussions after (\ref{fufu}) and after (\ref{jo6}),
and expanding $A$ over its eigenvalues and eigenvectors, we conclude {\it heuristically}
that for the convergence radius of $\sum_{k=2}^\infty \epf_k(n) z^k$ in
(\ref{retmon}) to be sufficiently larger than $1$, it is necessary to
have for the transfer-matrices $T(x)$ (using notations
(\ref{karamba_0}))
\BEA
\label{cote}
\lambda[T(x)]
\not
\approx \spe_1[T(x)], \quad
\lambda[T(x)]\not
\approx 1,
\EEA
i.e., closer is $\lambda[T(x)]$ to $\spe_1[T(x)]$ and or $\lambda[T(x)]$
to $1$, more terms are needed in the expansion (\ref{retmon}) for the
reliable estimate of the entropy.  Note that if
$\lambda[T(x)]=\spe_1[T(x)]>\spe_2[T(x)]$, the first relation in
(\ref{cote}) should be modified to $\lambda[T(x)] \not\approx
\spe_2[T(x)]$. We shall meet such examples below; see (\ref{tat21}) and
the discussion before it. 

Recall from (\ref{mkno}) that for calculating the entropy we need to
know $\xi(z,n)$ in the vicinity of $z=1$ and $n=1$. If the qualitative
conditions (\ref{cote}) are satisfied, we expect that the vicinity of
$z=1$ and $n=1$ is included in the convergence area. The convergence of
expansions similar to (\ref{retmon}) is discussed in
\cite{artuso,mainieri,aurell}.  In particular,
Refs.~\cite{artuso,mainieri} employ criteria similar to (\ref{cote}) and
test them numerically. 

In the context of expansion (\ref{retmon}) we should mention the results
devoted to analyticity properties of the top Lyapunov exponent
\cite{arnold,peres} and of the entropy for HMP \cite{han_markus}. In
particular, Ref.~\cite{han_markus} states that the entropy $h$ of HMP is
an analytic function of the Markov transition probabilities
(\ref{makarov}), provided that these probabilities are positive. At the
moment it is unclear for the present author how {\it in general} this
analyticity result can be linked to the expansion (\ref{retmon}).
However, we show below on concrete examples that the expansion
(\ref{retmon}) can be recast into an expansion over the Markov
transition probabilities (\ref{makarov}). 

\section{The simplest Aggregated Markov Process.}
\label{aggr}

\subsection{Definition.}

An Aggregated Markov Process (sometimes called a Markov source) is a
particular case of HMP, where the probabilities $\pi(x|s)$ in (\ref{1})
take only two values $0$ and $1$ \cite{ash,ephraim_review}. Thus it is
defined by the underlying Markov process $\S$ together with a
deterministic function $F(s_i)$ that takes the realizations of the
Markov process to those of the aggregated process:
$\X=(\X_1,\X_2,...)=(F(\S_1),F(\S_2),...)$. The function $F$ is not
one-to-one so that at least two realizations of $\S$ are lumped together
into one realization of $\X$. 

\begin{figure}%[hb]
\vspace{0.2cm}
\includegraphics[width=6cm]{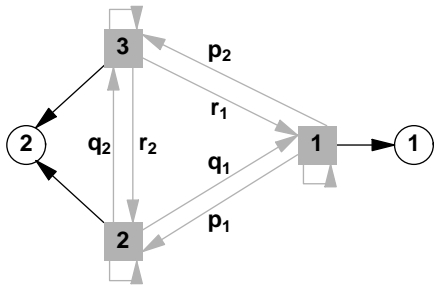}
\caption{
Schematic representation of the hidden Markov process defined by
(\ref{mack0}--\ref{mack2}). The gray squares and gray arrows indicate,
respectively, on the realization of the internal Markov process and transitions
between the realizations; see (\ref{mack1}). The circles and black arrows indicate on the
realizations of the observed process. The gray arrows are probabilistic;
the corresponding probabilities are indicated next to them. The black
arrow are deterministic; see (\ref{mack0}). 
}
\hfill
\label{fig_0}
\end{figure}

The simplest example is given by a Markov process
$\S=\{\S_0,\S_1,....\}$ with three realizations $\S_i=1,2,3$, such that,
e.g., the realizations $2$ and $3$ of $\S_i$ are not distinguished from each other
and correspond to one realization $2$ of the observed process $\X_i$ [see
Fig.~\ref{fig_0}]:
\BEA
\label{mack0}
&& F(1)=1,~~ F(2)=F(3)=2,\\
&&\pi(1|1)=1, ~~ \pi(1|2)=0, ~~ \pi(1|3)=0, \\
&&\pi(2|1)=0, ~~ \pi(2|2)=1, ~~ \pi(2|3)=1.
\EEA

The transition matrix of a general three-realization 
Markov process is [see Fig.~\ref{fig_0}]
\BEA
\label{mack1}
\P=
\left(\begin{array}{rrr}
1-p_1-p_2 & q_1~~~~ & r_1~~~~~ \\
p_1~~~~~  & 1-q_1-q_2  & r_2~~~~~  \\
p_2~~~~~ & q_2~~~~   & ~~1-r_1-r_2   \\
\end{array}\right),\qquad
|{\rm st}\rangle\propto
\left(\begin{array}{r}
q_1(r_1+r_2)+q_2r_1 \\
r_2(p_1+p_2)+p_1r_1 \\
p_2(q_1+q_2)+p_1q_2  \\
\end{array}\right)
\EEA
where all elements of $\P$ are positive, and where we presented the
stationary vector $|{\rm st}\rangle$ up to the overall normalization
\footnote{Note that some authors present the Markov transition matrices
$\P$ is such a way that the elements in each raw sum to one. This
amounts to transposition of (\ref{mack1}).  The representation
(\ref{mack1}) is perhaps more familiar to physicists.}. 

The process $\X=\{\X_1,\X_2,....\}$ has two realizations: $\X_i=1,2$. 
The corresponding transfer matrices read from (\ref{transfer})
\BEA
\label{mack2}
T(1)=
\left(\begin{array}{rrr}
1-p_1-p_2 & q_1 & r_1 \\
0~~~~~    & 0     & 0  \\
0~~~~~    & 0     & 0   \\
\end{array}\right),
\qquad
T(2)=
\left(\begin{array}{rrr}
0~~~~~~ & 0~~~~ & 0~~~~~ \\
p_1~~~~~  & 1-q_1-q_2  & r_2~~~~~  \\
p_2~~~~~ & q_2~~~~   & ~~1-r_1-r_2   \\
\end{array}\right).
\EEA
Note that the second (sub-dominant) eigenvalue of the transfer-matrix
product $\T(\bx)=\prod_{k=1}^N T(x_k)$ (with separate transfer-matrices
defined by (\ref{mack2})) is equal to zero, since this eigenvalue can be
presented as that of the matrix $T(1)A$, where $A$ is some $3\times 3$
matrix. The only exclusion, which has a non-zero sub-dominant eigenvalue,
is the realization of $\X$ that does not contain $1$ at all: $\T(2...2)=T^N(2)$. 

The considered HMP (\ref{mack0}--\ref{mack2}) belongs to the class of
HMP with unambiguous symbol, since the Markov realization $1$ is not
corrupted by the noise; see Fig.~\ref{fig_0}. For such HMP,
Ref.~\cite{han_markus} reports several results on the analytic features
of the entropy. 

\subsection{Unifilar process. }

Before studying in detail the HMP defined by (\ref{mack0}--\ref{mack2}),
let us mention one example of HMP, where the entropy can be calculated
directly \cite{ash,ephraim_review}.  This unifilar process is defined as
follows \cite{ash}: for each realization $s_i$ of the
Markov process $\S$ consider realizations $s_j$ with a strictly positive
transition probability $p(s_j|s_i)>0$.  Now require that the
realizations $F(s_j)$ of $\X_j$ are distinct. Thus given the realization
$s_i$ of $\S_1$, there is one to one correspondence between the
realizations of $(\X_1,\X_2,...)$ and those of $(\S_1,\S_2,...)$.  Write
the block-entropy of $\X$ as
\BEA
\label{kkk}
H(\X_N,...,\X_1)=H(\X_N,...,\X_1|\S_1)+H(\S_1)-H(\S_1|\X_1,...,\X_N),
\EEA
where $H({\cal A}|{\cal B})\equiv -\sum_{a,b}{\rm Pr}(a,b)\ln {\rm
Pr}(a|b)$ is the conditional entropy of the stochastic variable ${\cal
A}$ given ${\cal B}$. Due to the definition of the unifilar process:
$H(\X_N,...,\X_1|\S_1)=H(\S_N,...,\S_2|\S_1)$. The latter is worked out
via the Markov feature:
\BEA
&& H(\S_N,...,\S_2|\S_1)=(N-1)h_{\rm markov},\\
\label{boruch}
\label{ban}
&& h_{\rm markov}=-\sum_{k,l} p_{\rm st}(k) p(l|k)\ln p(l|k),
\EEA
where $p_{\rm st}(k)$ is the stationary Markov probability defined in (\ref{graal}),
and where $p(l|k)$ are the Markov transition probabilities from (\ref{makarov}). 
Since $H(\S_1)$ and $H(\S_1|\X_1,...,\X_N)$ in (\ref{kkk})
are finite in the limit $N\to\infty$, the entropy $h(\X)$ of the unifilar process
reduces to that of the underlying Markov process
$h_{\rm markov}$ \cite{ash}.

Note that any finite-order Markov process (conventionally assuming that
the usual Markov process is of first order) can be presented as a
unifilar process. There are, however, unifilar processes that do not
reduce to any finite-order Markov process \cite{ash}\, \footnote{The
example of such a process given in \cite{ash} is not minimal. The
minimal example is given by four-realization Markov process with
non-zero transition probabilities $p(4|1)$, $p(3|4)$, $p(2|3)$,
$p(1|2)$, $p(1|1)$, $p(2|2)$, $p(3|3)$ and $p(4|4)$ (all other
transition probabilities are zero), and two realizations of $\X_i$ such
that $F(1)=F(3)=1$, $F(2)=F(4)=2$. The unifilar process $\X$ does not
reduce to a finite-order Markov process, since, e.g., there are two
different mechanisms of producing the sequence $1...1$. This means that
$P(1|111)$ is not equal to $P(1|11)$, {\it etc}. }.
The main problem in identifying unifilar processes is that even if $\X$
is not unifilar for given $\S$, it can be still unifilar with respect to
another Markov process $\S'$ (see section \ref{dard} below for the
simplest example).  This makes especially difficult the recognition of
unifilar processes that do not reduce to any finite-order Markov
process. 

\subsection{Particular cases. }
\label{dard}

We now return to the HMP (\ref{mack0}--\ref{mack2}) and 
discuss some of its particular cases.

{\bf 1.} For $q_2=r_2$ and $q_1=r_1$ all
the terms $\varphi_k$ with $k\geq 3$ in the expansion (\ref{retmon})  
are zero. One can check that for this
case the observed process $\X$ is by itself Markov. 

{\bf 2.} For $(1-q_1-q_2)(1-r_1-r_2)=q_2r_2$, one can check that
$\phi_k=0$ for $k\geq 4$. Now the process $\X$ is the second-order 
Markov: $P(x_k|x_{k-1}, x_{k-2}, x_{k-3}) = P(x_k|x_{k-1}, x_{k-2})$.

Thus at least for these two cases the calculation of the entropy is straightforward.

The above two facts tend to clarify the meaning of the expansion
(\ref{retmon}).  It is tempting to suggest that if the expansion
(\ref{retmon}) is cut precisely at a positive integer $K> 2$, i.e.,
$\varphi_{k\geq K}=0$, then the corresponding process $\X$ is $K-2$-order
Markovian. If true, this will give convenient conditions for deciding on
the finite-order Markov feature, and will mean that the successive terms in
(\ref{retmon}) are in fact approximations the HMP via finite-order Markov
processes.

\subsection{Upper and lower bounds for the entropy.}

Before presenting the main results of this section, let us 
recall that the entropy of any (stationary) HMP satisfies
the following inequalities \cite{cover_thomas}
\footnote{Eq.~(\ref{mukik}) is a particular case of a slightly more general inequality
\cite{cover_thomas,birch}.
For our purely illustrative purposes (\ref{mukik}) is sufficient.}:
\BEA
\label{mukik}
H(\X_2|\S_1)
\leq
h\leq H(\X_2|\X_1)\equiv H(2)-H(1),
\EEA
where 
$H({\cal A}|{\cal B})=-\sum_{a,b} {\rm Pr}({\cal A} = a, {\cal B} = b)
\ln {\rm Pr}({\cal A} = a| {\cal B} = b)$ and $H(N)$ are, respectively, 
the conditional entropy and the block entropy
defined in (\ref{block}).  Employing (\ref{1}, \ref{transfer}) we deduce
\BEA
\label{orto}
{\rm Pr}(\X_2 = x| \S_1 = s)=\sum_{s'=1}^L T_{s'\, s}(x).  
\EEA
This equation
together with the stationary probability (\ref{mack1}) of the Markov process
is sufficient for calculating $H(\X_2|\S_1)$ for the HMP (\ref{mack0}, \ref{mack2}):
\BEA
H(\X_2|\S_1)=p_{\rm st}(1)\chi(p_1+p_2)+p_{\rm st}(2)\chi(q_1)+p_{\rm st}(3)\chi(r_1),\\
\chi(p)\equiv -p\ln p-(1-p)\ln(1-p).
\EEA
The upper bound $H(\X_2|\X_1)$ is calculated directly from (\ref{mubarak},
\ref{laokon}, \ref{block}).

\subsection{Generating function and entropy: exact results.}
\label{exo}

For a particular four-parametric class of HMP (\ref{mack0}--\ref{mack2})
we were able to sum exactly the expansion (\ref{retmon}) \footnote{This 
was done by hands, checking the separate terms of the expansion (\ref{castro}).}.
This class is characterized by the condition that
the two leading eigenvalues of the transfer-matrix $T(2)$ in (\ref{mack2}) have
equal absolute values [the third eigenvalue is equal to zero]: 
\BEA
\label{berd}
\lambda[T(2)]
= \lambda_1[T(2)].
\EEA
A direct
inspection shows that this condition amounts to two possible forms 
(\ref{moroz1}) and (\ref{moroz2}) of the
transition matrix $\P$. These two cases are studied below. 

\subsubsection{First case.}

\begin{figure}%[hb]
\vspace{0.2cm}
\includegraphics[width=7cm]{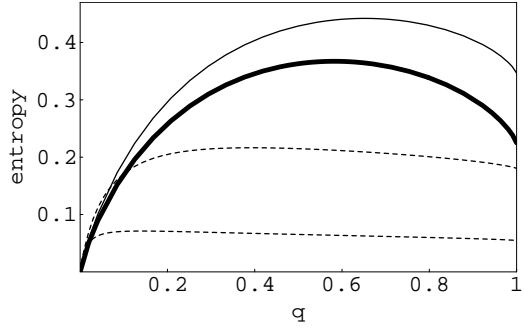}
\caption{Entropy (\ref{mukiko}) of HMP (\ref{mack1}, \ref{mack2}, \ref{moroz1})
versus $q=q_2$ for $p_2=q_1=0$. Normal line: $p_1=0.5$. Thick line: $p_1=0.75$. 
Upper dashed line: $p_1=0.05$. Lower dashed line: $p_1=0.01$. 
It is seen that for a small value of $p_1$,
the entropy $h$ is nearly constant for a range of $q=q_2$.
}
\hfill
\label{fig_exact}
\end{figure}

For this first case the transition matrix is obtained from (\ref{mack2})
under \footnote{Or, alternatively, via $q_2=0$ and $q_1=r_1+r_2$.
This, however, does not amount to anything new as compared to (\ref{moroz1}). }
\BEA
\label{moroz00}
r_2=0 \quad {\rm and} \quad r_1=q_1+q_2.
\EEA
This leads from (\ref{mack1}) to the transition matrix
\BEA
\P=
\left(\begin{array}{rrr}
1-p_1-p_2 & q_1~~~~ & ~~q_1+q_2~~ \\
p_1~~~~~  & 1-q_1-q_2  & 0~~~~~  \\
p_2~~~~~ & q_2~~~~   & ~~1-q_1-q_2   \\
\end{array}\right).
\label{moroz1}
\EEA
It is seen that the realization $\{\S_{k+1}=2,\,\S_{k}=3\}$ for the Markov process is prohibited. For the HMP
there are no prohibited sequences. 

The inverse zeta-function reads from (\ref{retmon}):
\BEA
\label{tat21}
\xi(z,n)&=&1-[\,(1-p_1-p_2)^n+(1-q_1-q_2)^n\,]\, z \nonumber\\
&+&[\,(1-p_1-p_2)^n (1-q_1-q_2)^n
-(p_1q_1+p_2(q_1+q_2)\,)^n
\,] z^2\nonumber\\
&+& z^3 [p_1q_2(q_1+q_2)]^n\left[\,
\Phi(y,-n,b)-\Phi(y,-n,b+1)
\,\right],
\label{tat2}
\EEA
where we defined 
\BEA
\label{koma}
&& b\equiv (1-q_1-q_2)\frac{p_2(q_1+q_2)+p_1q_1}{p_1q_2(q_1+q_2)},\\
&& y\equiv (1-q_1-q_2)^n z,
\EEA
and where $\Phi(y,-n,b)$ is the Lerch $\Phi$-function:
\BEA
\Phi(y,-n,b)=\sum_{k=0}^\infty (k+b)^n y^k . 
\EEA
In this representation, which led to (\ref{tat2}), the sum converges
for $|y|<1$ or for $z<(1-q_1-q_2)^{-n}\geq 1$. 
The convergence radius tends to one for $q_1+q_2\to 0$, or, equivalently, for
$\lambda[T(2)] \to 1 $; see (\ref{mack2}). 
This violates the second qualitative condition in (\ref{cote}).

Using (\ref{mkno}) we get from (\ref{tat2}) for the entropy:
\BEA
&&h=-\frac{1}{p_1+p_2+q_1+q_2+\frac{p_1q_2}{q_1+q_2}}\{\, \nonumber\\
&&(1-p_1-p_2)(q_1+q_2)\ln(1-p_1-p_2)+(1-q_1-q_2)(p_1+p_2+\frac{p_1q_2}{q_1+q_2})\ln(1-q_1-q_2)\nonumber\\
&&+p_1q_2\ln [\,p_1q_2(q_1+q_2)\,]
+[\,(p_1+p_2)q_1+p_2q_2\,]\ln [\,(p_1+p_2)q_1+p_2q_2\,]
\nonumber\\
&&+p_1q_2(q_1+q_2)\left[\,
\Phi'_{[2]}(1-q_1-q_2,-1,b) - \Phi'_{[2]}(1-q_1-q_2,-1,b+1)
\,\right]
\},
\label{mukiko}
\EEA
where $b$ is defined in (\ref{koma}), 
and where 
\BEA
\Phi'_{[2]}(y,-1,b)=
\sum_{k=0}^\infty \ln\left[\frac{1}{k+b}  \right](k+b) y^k .
\EEA 

The behavior of $h$ is illustrated in Fig.~\ref{fig_exact} for 
particular values of $p_1$, $p_2$, $q_1$ and $q_2$. Table~\ref{tab_0000}
compares the exact expression (\ref{mukiko}) with the upper and lower bounds
(\ref{mukik}).

The analytic features of $h$ given by (\ref{mukiko}) as a function of
the Markov transition probabilities $p_1$, $p_2$, $q_1$ and $q_2$, agree
with the results obtained in \cite{han_markus}. In particular, note that
for $p_1+p_2\to 1$ the entropy $h$ becomes non-analytic due to the term $\propto
(1-p_1-p_2)\ln (1-p_1-p_2)$.

\begin{table}
\caption{For two set of parameters of 
the HMP (\ref{mack0}, \ref{moroz00}, \ref{moroz1}) we present the 
exact value of entropy $h$ obtained from (\ref{mukiko}), the lower bound 
$H(\X_2|\S_1)$, and the upper bound $H(\X_2|\X_1)$; see (\ref{mukik}). 
%%The parameters $p_1$, $p_2$, $q$ and $r$ are tuned such that 
%%$H(\X_2|\S_1)$ and $H(\X_2|\X_1)$ provide rather tight bounds on $h$.
}
\begin{tabular}{|c||c|c|c|}
\hline
             & $h$ & $H(\X_2|\S_1)$ & $H(\X_2|\X_1)$ \\
\hline\hline
$p_1=0.75$ &          &          &          \\
$p_2=0.10$ & 0.569580 & 0.557243 & 0.572373  \\
$q_1=0.25$  &          &          &           \\
$q_2=0.20$   &          &           &          \\
\hline \hline
$p_1=0.30$ &          &          &           \\
$p_2=0.20$ & 0.684796 & 0.682486 & 0.684843  \\
$q_1=0.55$   &          &          &           \\
$q_2=0.10$   &          &           &          \\
\hline \hline
\end{tabular}
\label{tab_0000}
\end{table}

\begin{figure}%[hb]
\vspace{0.2cm}
\includegraphics[width=7cm]{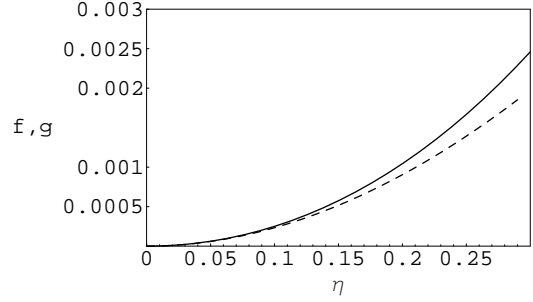}
\caption{The rate functions $f(\eta)$ and $g(\eta)$ defined 
by (\ref{gusi_f}) and (\ref{gusi_g}), respectively 
for the HMP given by (\ref{mack2}, \ref{mack99}, \ref{m_bao}). 
Normal line : $g(\eta)$. Dashed line : $f(\eta)$. 
For the parameters in (\ref{mack99}) we take: 
$p_1=0.2$, $p_2=0.3$, $q=0.05$, and $r=0.01$.
For these values the entropy (\ref{mukik1}) is $h=0.166671$.
}
\hfill
\label{fig_2}
\end{figure}

\begin{figure}%[hb]
\vspace{0.2cm}
\includegraphics[width=7cm]{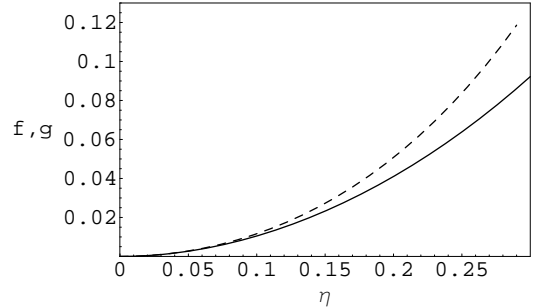}
\caption{The same as in Fig.~\ref{fig_2} but with $q=0.1$ and $r=0.4$. For these values
the entropy (\ref{mukik1}) is $h=0.619519$, which is larger than the entropy in Fig.~\ref{fig_2}.
}
\hfill
\label{fig_3}
\end{figure}

\subsubsection{Second case.}

The second possibility of satisfying (\ref{berd}) is given by
\BEA
q_1+q_2=1 \quad {\rm and} \quad
r_1+r_2=1,
\EEA
\BEA
\P=\left(\begin{array}{rrr}
1-p_1-p_2  & q~~~   & r~~ \\
p_1~~~  & 0~~~         & 1-r    \\
p_2~~~  & ~1-q~~     & 0~~   \\
\end{array}\right).
\label{moroz2}
\label{mack99}
\EEA
The realizations of the corresponding Markov process do not contain 
$\{\S_{k+1}=2,\,\S_{k}=2\}$ and $\{\S_{k+1}=3,\,\S_{k}=3\}$.
Again, the realizations of the HMP do not have any prohibited sequence. 

The inverse zeta-function reads from (\ref{retmon})
\BEA
\label{m_bao}
\xi(z,n)&=&1-\left[\,(1-p_1-p_2)^n+(1-q)^{n/2}(1-r)^{n/2}\,\right]\, z+
\left[\,
-(p_1q+p_2r)^n + (1-p_1-p_2)^n (1-q)^{n/2}(1-r)^{n/2}
\,\right] z^2\nonumber\\
&+& \frac{z^3}{1+z (1-q)^{n/2}(1-r)^{n/2} }
\left[\,
(p_1q+p_2r)^n (1-q)^{n/2}(1-r)^{n/2} -(\, p_1r(1-q) + p_2 q (1-r)   \,)^n
\,\right].
\label{m_bao1}
\EEA
The series that led to (\ref{m_bao}) converges 
for $|z|<(1-q)^{-n/2}(1-r)^{-n/2}$. Again the 
convergence radius going to one violates the 
second qualitative condition in (\ref{cote}).

Eqs.~(\ref{ad_din}, \ref{m_bao}) imply for the source entropy:
\BEA
h=-\frac{1}{\,\,2(p_1+p_2)+q(1-p_1)+r(1-p_2)-qr\,\,}\left\{\,
[\,q(1-r)+r\,](1-p_1-p_2)\ln (1-p_1-p_2)\right.\nonumber\\ \left.
+(p_1+p_2)\,(1-q)\,(1-r)\,\ln [(1-q)(1-r)]
+(p_1q+p_2r)\ln [p_1q+p_2r]
\right.\nonumber\\ \left.
+[p_2q(1-r)+p_1(1-q)r] \ln [p_2q(1-r)+p_1(1-q)r]
\,\right\}.
\label{mukik1}
\EEA

Applying the general definition (\ref{boruch}) of the Markov entropy 
to the particular case (\ref{mack1}) we get for the Markov entropy
\BEA
h_{\rm markov}&=&-\frac{1}{\,\,2(p_1+p_2)+q(1-p_1)+r(1-p_2)-qr\,\,}\,\{\,
\nonumber\\ 
&&[\,q(1-r)+r\,][\, (1-p_1-p_2)\ln (1-p_1-p_2)+ p_1\ln p_1 + p_2\ln p_2\, ]
\nonumber\\  
&&[(1-r)(p_1+p_2)+p_1 r]\, [q\ln q+(1-q)\ln (1-q)]
\nonumber\\ 
&&[p_2+p_1(1-q)]\, [r\ln r+(1-r)\ln (1-r)]
\,\}.
\label{mukik2}
\EEA

Comparing (\ref{mukik1}, \ref{mukik2}) one can check [e.g., numerically]
that $h_{\rm markov}>h$, as should be, since lumping several states
together decreases the entropy. Table \ref{tab_0} compares the exact value (\ref{mukik1}) for the
entropy with the upper and lower bounds (\ref{mukik}).

\begin{table}
\caption{For two set of parameters of 
the HMP (\ref{mack1}, \ref{mack2}, \ref{moroz2}) we present the 
exact value of entropy $h$ obtained from (\ref{mukik1}), the lower bound 
$H(\X_2|\S_1)$, and the upper bound $H(\X_2|\X_1)$; see (\ref{mukik}). 
The parameters $p_1$, $p_2$, $q$ and $r$ are tuned such that 
$H(\X_2|\S_1)$ and $H(\X_2|\X_1)$ provide rather tight bounds on $h$.
}
\begin{tabular}{|c||c|c|c|}
\hline
             & $h$ & $H(\X_2|\S_1)$ & $H(\X_2|\X_1)$ \\
\hline\hline
$p_1=0.1$ &          &          &          \\
$p_2=0.1$ & 0.528531 & 0.525571 & 0.528534  \\
$q=0.2$  &          &          &           \\
$r=0.3$   &          &           &          \\
\hline \hline
$p_1=0.2$ &          &          &           \\
$p_2=0.2$ & 0.659897 & 0.656974 & 0.659901  \\
$q=0.3$   &          &          &           \\
$r=0.4$   &          &           &          \\
\hline \hline
\end{tabular}
\label{tab_0}
\end{table}

\subsubsection{Rate functions for large deviations.}
\label{dobre}

Recall that the rate function $f(\eta)$ ($g(\eta)$) defined in section
\ref{daos}, describe the weight of atypical sequences with the
probability smaller (larger) than the typical sequence probability
$e^{-Nh}$. The positive parameter $\eta$ defines the amount of this
smallness (largeness); see (\ref{gusi_f}) and (\ref{gusi_g}). 

The calculation of $f(\eta)$ and $g(\eta)$ for the considered
HMP model (\ref{moroz2}, \ref{mack2}) is straightforward. One finds out the
zero of the $\xi$-function given by (\ref{m_bao1}). This will define,
via (\ref{khrych}), the moment-generating function $\Lambda(n)$. If
there are several zeros of $\xi(z,n)$ as a function of $z$, we select
the one that goes to $z=1$ for $n\to 1$. Then $f(\eta)$ and $g(\eta)$
are calculated from their definitions (\ref{gusi_f}) and (\ref{gusi_g}). 

The behavior of $f(\eta)$ and $g(\eta)$ as functions of $\eta$ is
presented in Figs. \ref{fig_2} and \ref{fig_3}. For each figure we take
different sets of parameters $p_1$, $p_2$, $q$ and $r$; see
(\ref{moroz2}) for their definition.  To make this difference explicit
let us denote $f_3(\eta)$, $g_3(\eta)$ and $f_4(\eta)$, $g_4(\eta)$ for
Fig. \ref{fig_2} and Fig. \ref{fig_3}, respectively. 

Now let us observe that 
\BEA
\label{kaa1}
f_3(\eta)<f_4(\eta), \qquad g_3(\eta)<g_4(\eta),
\EEA
\BEA
g_3(\eta)>f_3(\eta), \qquad g_4(\eta)<f_4(\eta).
\label{kaa2}
\EEA

For explaining these inequalities we note that for the parameters
of Fig.~\ref{fig_2} the entropy is smaller than $h$ in
Fig.~\ref{fig_3}:
\BEA
\label{kaa3}
h_3<h_4,
\EEA
which means that the typical set $\Omega_N^*$ for Fig.~\ref{fig_3}
contains more sequences, so there remains less of them outside, which
may explain (\ref{kaa1}).  For the same reason (\ref{kaa3}), the
probability of each typical sequence is higher for the parameters in
Fig.~\ref{fig_2}.  Thus for the parameters presented in Fig.~\ref{fig_2}
more high-probability sequences are included in the corresponding typical set
$\Omega_N^*$. This may explain (\ref{kaa2}). 

In further numerical checkings it was noted that the above relation
between (\ref{kaa1}) and (\ref{kaa2}) from one side, and (\ref{kaa3})
from another side, seems to be much more general than these particular
examples. 

\section{Binary symmetric Hidden Markov Process.}
\label{bshmp}

\subsection{Definition and symmetries.}

This is another popular (and simple to define) example of HMP.
Now the Markov process has two states $1$ and $2$.
The realizations of the observed (Hidden Markov) process also take two values $1$ and $2$. 
The internal Markov process is driven by the conditional probability
\BEA
\label{o1}
&&\P=
\left(\begin{array}{rr}
p(1|1)~~ & ~~p(1|2)~~ \\
\\
p(2|1)~~ & ~~p(2|2)~~ \\
\end{array}\right)
=
\left(\begin{array}{rr}
1-\p & ~~\p~~ \\
\\
\p~~ & ~~1-\p \\
\end{array}\right).
\EEA
The stationary probability for this Markov process is found
via (\ref{graal}): $p_{\rm st}(1) =p_{\rm st}(2)=\frac{1}{2}$.

The probabilities for the observations $1$ or $2$ given the internal state read
\BEA
\label{o2}
&&\pi(\x_i|s_i) =
\left(\begin{array}{rr}
\pi(1|1)~~ & ~~\pi(1|2)~~ \\
\\
\pi(2|1)~~ & ~~\pi(2|2)~~ \\
\end{array}\right)
=\left(\begin{array}{rr}
1-\e & ~~\e~~ \\
\\
\e~~ & ~~1-\e \\
\end{array}\right),
\EEA
where $\e$ is the error probability during the observation.

For the transfer matrices we have:
\BEA
\label{do1}
T(1)=
\left(\begin{array}{rr}
\e(1-\p)~~ & ~~\e\p~~~~~~~~~ \\
\\
(1-\e)\p~~ & ~~(1-\e)(1-\p)~~ \\
\end{array}\right), \qquad
T(2)=
\left(\begin{array}{rr}
(1-\e)(1-\p)~~ & ~~(1-\e)\p~~ \\
\\
\e\p~~~~~~~~~ & ~~\e(1-\p)~~ \\
\end{array}\right).
\label{bakunin}
\EEA
$T(2)$ is obtained from $T(1)$ via $\e\to 1-\e$.

The following symmetry features are deduced directly from (\ref{o1}--\ref{bakunin}):

(1) For any $N$ the probability $P(x_N,\ldots,x_1; q, \e)$ of the binary
symmetric HMP is invariant with respect to $\e\to 1-\e$:
$P(x_N,\ldots,x_1; q, \e)=P(x_N,\ldots,x_1; q, 1-\e)$. 

(2) The probability $P(x_N,\ldots,x_1; q, \e)$ is invariant with respect to the full "inversion"
of the realization $(x_N,\ldots,x_1)$, e.g. $P(1,2,1,1; q, \e)=P(2,1,2,2; q, \e)$.

(3) In general, the probability $P(x_N,\ldots,x_1; q,\e)$ is not
invariant with respect to $q\to 1-q$, e.g.,
$P(1,2;q,\e)-P(1,2;1-q,\e)=\frac{1}{2}(1-2\e)(2q-1)$. However, for each
given realization $(x_N,\ldots,x_1)$ one can find another unique
realization $(\bar{x}_N,\ldots,\bar{x}_1)$ such that $P(x_N,\ldots,x_1;
q,\e)=P(\bar{x}_N,\ldots,\bar{x}_1;1-q,\e)$.  The logics of relating
$(x_N,\ldots,x_1)$ to $(\bar{x}_N,\ldots,\bar{x}_1)$ should be clear
from the following example: if $(x_4,\ldots,x_1)=(1,2,2,1)$, then
$(\bar{x}_4,\ldots,\bar{x}_1)=(2,2,1,1)$. In more detail, $\bar{x}_4=2$
is defined to be different from $x_4=1$, and once $x_3=2$ is different
from $x_4=1$, $\bar{x}_3=2$ does not differ from $\bar{x}_4=2$, {\it
etc}. It should be clear (e.g., by induction) that for a given
$(x_N,\ldots,x_1)$, $(\bar{x}_N,\ldots,\bar{x}_1)$ is indeed unique. 

This feature means, in particular, that the entropy $h$ of the binary
symmetric HMP|being according to (\ref{block}, \ref{hopo}) a symmetric
function of all probabilities $P(x_N,\ldots,x_1)$|is invariant with
respect to $q\to 1-q$: $h(q,\e)=h(1-q,\e)$, in addition to being invariant
with respect to $\e\to 1-\e$.

(4) In general, the probabilities $P(x_N,\ldots,x_1)$ are not invariant with respect to 
a cyclic interchange of the realizations, e.g., $P(1,2,1;q,\e)-P(1,1,2;q,\e)=\frac{1}{2}(1-2\e)^2q(2q-1)$.

\begin{figure}%[hb]
\vspace{0.2cm}
\includegraphics[width=7cm]{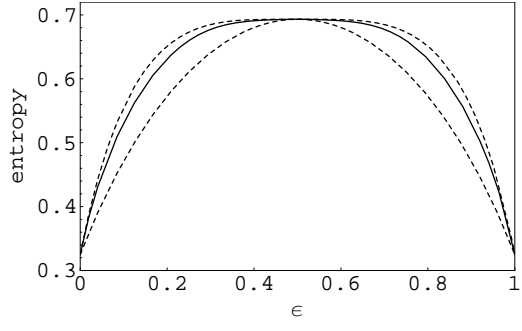}
\caption{Entropy of the binary hidden Markov chain (normal line) versus the error probability
$\e$ for $q=0.1$. Dashed lines: upper and lower bounds for the entropy as given by 
(\ref{mukik}). The entropy is calculated from (\ref{retmon}, \ref{mkno}) approximating the
infinite sum in (\ref{retmon}) by a poynomial or the order $13$.
}
\hfill
\label{fig_1}
\end{figure}

For the considered binary symmetric HMP we did not find any exactly
solvable situation. Thus, we employed (\ref{retmon}) and calculated
$\xi(z,n)$ by approximating the infinite sum in the RHS of
(\ref{retmon}) via a polynom of order $K$: $\sum_{k=2}^K \epf_k(n)
z^k$ \, \footnote{The terms in this expansion can perhaps be re-arranged so as to facilitate 
the convergence. Since in the present paper the numerical calclations serve mainly illustrative 
purposes, we shall not dwell into this aspect.}. 
This approximation was suggested in \cite{artuso} and it is based
on the fact that the sum supposed to converge exponentially at least in
the vicinity of $z= 1$ and $n= 1$. This is what we saw for the
exactly solvable situations (\ref{tat21}) and (\ref{m_bao}). 
The qualitative criterion for the exponential converges
was suggested in \cite{artuso,mainieri} and was discussed by us around (\ref{cote}).
Since both transfer-matrices in (\ref{do1}) have the 
same eigenvalues 
\BEA
\frac{1}{2}\left[\,
1-q\pm\sqrt{q^2+(1-2q)(1-2\epsilon)^2}
\,\right],
\label{gamadril}
\EEA
for the studied binary symmetric HMP there are several cases, where the
[qualitative] conditions (\ref{cote}) are violated: {\it i)} $q\to 0$
and $\e\to\frac{1}{2}$; {\it ii)} $q\to 1$; {\it iii)} $q\to 0$ and
$\e\to 0$. In these three cases we expect that that approximating
$\xi(z,n)$ by $\sum_{k=2}^K \epf_k(n) z^k$ will not be feasible, since
large values of $K$ will be required to achieve a reasonably high
precision.  Fig.~\ref{fig_1} and Table \ref{tab_00} present the results
for the entropy obtained in the above approximate way and compare them
with the upper and lower bounds, as given by (\ref{mukik}). 

\begin{table}
\caption{For two sets of the parameters $q$ and $\epsilon$ of 
the binary symmetric HMP (\ref{o1}, \ref{o2}, \ref{do1}) we present the 
entropy $h$ obtained by approximating (\ref{retmon}) via a polynomial or order $2$, $13$ and $12$,
respectively. These values are denoted by $h_2$, $h_{13}$ and $h_{12}$. We
compare $h_{k}$ with the lower bound
$H(\X_2|\S_1)$, and the upper bound $H(\X_2|\X_1)$; see (\ref{mukik}). 
It is seen that the relative difference $\frac{h_{13}-h_{2}}{h_{13}}$ is
not larger than $0.02$.
}
\begin{tabular}{|c||c|c|c|c|c|}
\hline
             & $h_{2}$    & $h_{13}$ & $h_{12}$ & $H(\X_2|\S_1)$ & $H(\X_2|\X_1)$ \\
\hline\hline
$q=0.2$ &    &      &          &          & \\
$\e=0.45$  & 0.687811     & 0.693108 & 0.693100 & 0.691346 & 0.693129 \\
\hline \hline
$q=0.25$ &  &        &          &          & \\
$\e=0.4$ & 0.681322   &  0.692884 & 0.692881 & 0.688139 & 0.692947 \\
\hline \hline
\end{tabular}
\label{tab_00}
\end{table}

\subsection{Small-noise limit.}

For $\e=\frac{1}{2}$ or for $q=\frac{1}{2}$ the process becomes
memory-less: $P(x_1,...,x_N)=P(x_1)...P(x_N)$. Here all the functions $\varphi_k$
in (\ref{retmon}) are equal to zero. Another particular case
is the limit $\e\to 0$ (no noise), where the hidden Markov process
degenerates into the original Markov process. It is straightforward to
check that in (\ref{retmon}) for the entropy only the term $\phi_2$ is
different from zero, while $\phi_k=0$ for $k\geq 3$.  This produces the
well-known expression (\ref{ban}) for the entropy of a Markov process. 

Let us work out the vicinity of $\e=0$, assuming that $\e$ is small (quasi-Markov situation).
One can check that 
\BEA
\varphi_{k}={\cal O}(\e^{k-2}) \qquad {\rm for} \qquad k\geq 3.
\EEA
Thus for finding the entropy and the generating function within the
order ${\cal O}(\e^2)$, we need to expand $\varphi_k$ with $k=1,2,3,4$
over $\e$ and select all the terms of order ${\cal O}(\e)$ and ${\cal
O}(\e^2)$. We write down explicitly the approximation of $\xi(z,n)$ via
the polynom of order $4$ (higher-order terms $\varphi_{k\geq 5}$ are not
needed, since they do not contribute to the order ${\cal O}(\e^2)$):
\BEA
\label{bobo}
\xi(z,n)=1+z\,\varphi_1(n)+z^2\varphi_2(n)+z^3\varphi_3(n)+z^4\varphi_4(n)+
{\cal O}(z^3). 
\EEA
Using (\ref{gamadril}) and (\ref{jo2}--\ref{jo4.2}) 
we get after straightforward algebraic 
calculations (taking for simplicity $q<\frac{1}{2}$)
\BEA
\varphi_1(n)&=&-2\,{\left( 1 - q \right) }^n \nonumber\\
            &+& 2\,\e\,n\,{\left( 1 - q \right) }^{n-2}\,\left( 1 - 2\,q \right) \nonumber \\
            &-& 
  \e^2\,n\,{\left( 1 - q \right) }^{n-4}\,\left( 1 - 2\,q \right) \,
   \left\{ \, (1-2q)(n-1-q)+q \,\right\}+{\cal O}(\e^3), 
\label{ra1}
\EEA
\BEA
\varphi_2(n)&=&
{\left( 1 - q \right) }^{2\,n} - q^{2\,n} \nonumber\\
&-& 2\,\e\,n\,\left( 1 - 2\,q \right) \,
   \left[ {\left( 1 - q \right) }^{2\,\left( n-1 \right) } + q^{2\,(n-1)} \right] \nonumber \\
  &-&\e^2\,n\,\left( 1 - 2\,q \right) \,\left[ q^{ 2\,(n-2)}\,
      \left\{\,  (1-2q)(q+2n-3)-q\,\right\}\right.\nonumber\\
&&~~~~~~~~~~~~~~~~~\left.+
     ( 1 - q )^{2( n-2 ) }\,
      \left\{\, (1-2q)(q+1-2n)-q\,\right\} \right]+{\cal O}(\e^3), 
\label{ra2}
\EEA
\BEA
\varphi_3(n)&=&2\e\,n\, (1-2q)^2\, (1-q)^{n-2}\, q^{2(n-1)}\nonumber\\
&-&\e^2\,n\,(1-2q)^2\,(1-q)^{n-4}\,q^{2(n-2)}\,\left[\,
5-3n+4q(3n-5)+2q^2(16-7n)\right.\nonumber\\
&&~~~~~~~~~~~~~~~~~~~~~~~~~~~~~~~~~~~~~~~~~~
\left.+4q^3(n-6)+10q^4
\,\right]+{\cal O}(\e^3), 
\label{ra3}
\EEA
\BEA
\label{ra4}
\varphi_4(n)=
\e^2 n\, (1-2q)^3\, (1-q)^{2(n-2)}\, q^{2(n-2)}\,\left[\,
2-4q(1-q)-n(1-2q)
\,\right]+{\cal O}(\e^3).
\EEA
Note that all $\e$ corrections nullify for $q=\frac{1}{2}$, once in this limit we
should get a memory-less process.
These equations produce for the entropy from (\ref{bobo}, \ref{mkno}):
\BEA
\label{kaz1}
h&=&-(1-q)\ln (1-q)-q\ln q\\
\label{kaz2}
&-&2\e\,(1-2q)\,\ln\left(\frac{1-q}{q}\right)\\
\label{kaz3}
&-&2\e^2\,(1-2q)\left[
\, \ln\left(\frac{1-q}{q}\right)+\frac{1-2q}{4(1-q)^2\,q^2}
\,\right]+{\cal O}(\e^3).
\EEA
Eq.~(\ref{kaz1}) is just the Markov entropy (\ref{boruch}) obtained in
the limit $\e=0$.  Eqs.~(\ref{kaz2}) is the first correction to the
Markov situation; it is obtained in \cite{jacquet,weissman}.  The second
correction (\ref{kaz3}) is reported in \cite{zuk_jsp}. The authors of \cite{zuk_jsp}
also obtain the higher-order corrections employing the mapping of the
binary symmetric HMP to the one-dimensional Ising model. These
higher-order correction can be also obtained within the present method.
Thus we demonstrated that the small-noise (quasi-Markov) situation can be
adequately explored with the present method. 

In addition we obtain the small-noise expressions (\ref{ra1}-\ref{ra4})
for the zeta-function.  This result is new and it allows to find the
moment-generating function, which contains more information than the
entropy, e.g., (\ref{bobo}--\ref{ra4}) can be used for approximating the
rate functions (\ref{gusi_f}) and (\ref{gusi_g}). In particular, for the
generating function we get from (\ref{khrych}) and
(\ref{ra1}--\ref{ra4})
\BEA
\Lambda(n)=q^n+(1-q)^n -\frac{\e n (1-2q)\left[\,(1-q)^{2n}q^2-(1-q)^{2}q^{2n}      \,\right]}
{q^2 (1-q)^2 \left[\,  (1-q)^n +q^n  \,\right]} + {\cal O}(\e^2).
\EEA

\section{Summary.}

In this paper we studied the entropy and the moment-generating function
of Hidden Markov Processes (HMP). The fact that these processes model
non-Markov memory is at the origin of their numerous applications, and,
simultaneously, the main reason of difficulties in characterizing their
entropy and the moment-generating function.  Recall that the entropy
gives the number of sequences in the typical set of the random process
\cite{cover_thomas,strat}; the typical set is the smallest set of
realizations with the overall probability close to one.  Alternatively,
the entropy is the uncertainty [per time-unit] of the process given its
long history. The generating function allows to estimate the [small]
probability of atypical sequences via the Chernoff bound and the rate
functions \cite{cover_thomas,strat}.  The entropy of HMP was studied via
upper and lower bounds \cite{cover_thomas,birch}, expansions over small
parameters \cite{zuk_jsp,zuk_aizenman,chigan}, and via expressing the
entropy as a solution of an integral equation
\cite{strat,blackwell,birch,jacquet,gold,weissman,egner}. 

Here we proposed to calculate the entropy and the moment-generating
function of HMP via the cycle expansion of the zeta-function, a method
adopted from the theory of dynamical systems \cite{artuso,mainieri,gaspar}. I
show that this method has two basic advantages. First, it produces exact
results, both for the entropy and the moment-generating function, for a
class of HMP. We did not so far got into any systematic way of searching
for the exact solutions within this method. The examples of exact
solutions presented in section \ref{exo} were obtained in the most
straightforward way.  Second, even if no exact
solution is found, the method offers an expansion for the entropy and
the moment-generating function via an exponentially convergent power
series \cite{artuso,mainieri,gaspar}.  Cutting off these expansions at some
finite order gives normally an improvable approximation for the sought
quantities, especially since there are qualitative estimates for the
convergence radius of the series. This was demonstrated in section \ref{bshmp}.

As a by-product of this study, we conjectured in section \ref{dard} on
tentative conditions under which HMP reduces to a finite-order Markov
process. These conditions compare favorably with those existing in
literature, see e.g. \cite{lolo}, and they deserve further exploration. 
We also conjectured relations (\ref{kaa1}--\ref{kaa3}) between the 
rate functions of the random process and its entropy.

\acknowledgements

I thank David Saakian for arousing my interest in this problem. 

The work was supported by Volkswagenstiftung grant
''Quantum Thermodynamics: Energy and information flow at nanoscale''.

\appendix

\section{Recollection of some facts about the eigen-representation 
versus singular value decomposition.}
\label{app_Weyl}
\label{silent_silent}

A matrix $A$ can be diagonalized if \cite{horn}
\BEA
A=V\,D\,V^{-1}, 
\EEA
where $D$ is a diagonal matrix, and where $V$ is an arbitrary
invertible matrix. Writing the eigen-resolution of $D$, 
$D=\sum_k \al_k|\al_k\rangle\langle \al_k|$, where 
$\langle \al_k|\al_n\rangle=\delta_{kn}$, one gets 
\BEA
A=\sum_k \al_k |R_k\rangle\langle L_k|,
\EEA
where $\al_k$ are the eigenvalues of $A$ (i.e., the solutions of
${\rm det\,}(A-\al\, 1)=0$), and where $|R_k\rangle$ and $|L_k\rangle$ are,
respectively, the right and left eigenvectors:
\BEA
A|R_k\rangle=\al_k|R_k\rangle, \qquad\langle L_k|A=\al_k\langle L_k|,\qquad
\langle L_k|R_n\rangle=\delta_{kn}.
\EEA
Note that in general $\langle L_k|L_n\rangle\not=\delta_{kn}$. The right
and left eigenvectors coincide for normal matrices $[A,A^\dagger]=0$ ($A$ commutes
with its complex conjugate). For those matrices $V$ is unitary. 

Not every matrix can be diagonalized, a necessary and sufficient
condition for this is that for each eigenvalue the algebraic degeneracy
(i.e., degeneracy of this eigenvalue as the root of the characteristic
polynom) coincides with the geometric degeneracy (the number of
eigenvectors corresponding to this eigenvalue; geometric degeneracy
cannot be larger than the algebraic one). Thus a sufficient condition
for a matrix to be diagonalizable is that its eigenvalues are not
degenerate. Here is a more general sufficient condition: Any matrix that
commutes with a matrix with non-degenerate eigenvalues, is
diagonalizable \cite{horn}. 

If for one eigenvalue $\al$ of $A$ the algebraic and geometric degeneracies are
equal (say to $m$), then
\BEA
A=V \left(\begin{array}{rr}
\al I_{m \times m}~~ & ~~0~~ \\
0~~ & ~~A'~~ \\
\end{array}\right)
V^{-1}, 
\EEA
where $I_{m \times m}$ is the $m \times m$ unit matrix.

An alternative representation for the matrix $A$ is given by the
singular value decomposition. Note that if ${\rm det}\,\,A\not =0$,
the matrix $A[A^\dagger A]^{-1/2}$ is unitary. Then it holds
\BEA
\label{hop}
A=U\,[A^\dagger A]^{1/2},
\EEA
where $U$ is unitary. Eq.~(\ref{hop}) holds also for ${\rm det}\,A=0$ via the
continuity. Going to the eigen-resolution of the hermitian matrix $A^\dagger A$,
we see that for {\it any} matrix $A$ there is a singular value decomposition:
\BEA
&& A=\sum_k \sigma_k |u_k\rangle\langle v_k|, \\
&& A|v_k\rangle=\sigma_k |u_k\rangle, \qquad  \langle v_k|v_n\rangle=\delta_{kn}  \\
&& \langle u_k|A =\sigma_k \langle v_k|, \qquad  \langle u_k|u_n\rangle=\delta_{kn},
\EEA
where $\sigma_k$ (singular values of $A$) 
is the common eigenvalue spectrum of $\sqrt{AA^\dagger}$ and
$\sqrt{A^\dagger A}$.

For a given diagonalizable matrix $A$, its singular value decomposition is 
related to the eigen-resolution via \cite{horn}
\BEA
\langle v_n|R_k\rangle \sigma_n=\al_k  \langle u_n|R_k\rangle, \\
\langle u_n|L_k\rangle \sigma_n=\al_k^*  \langle v_n|L_k\rangle.
\EEA

The matrix $A$ is normal if and only if $|\al_k|=\sigma_k$.  (I did not
find any standard reference on the fact that $|\al_k|=\sigma_k$ leads to
normality; the proof I got myself is too tedious to be presented here). 

Singular values and eigenvalues are related via the Weyl inequalities.
For a given matrix $A$, order the absolute values of its eigenvalues as
$\spe_0 \geq \spe_1\geq ... \geq \spe_n$,
and order its singular values as 
$\sigma_0 \geq \sigma_1\geq ... \geq \sigma_n$. 
The Weyl inequalities then read: 
\BEA
\label{we1}
&&\prod_{k=0}^m\sigma_{k}\geq
\prod_{k=0}^m\spe_{k},\quad
\prod_{k=0}^m\sigma_{n-k}\leq
\prod_{k=0}^m\spe_{n-k},\\
\label{we3}
&&
\sum_{k=0}^m\sigma_{i}^\rho\geq
\sum_{k=0}^m\spe_{i}^\rho, \qquad  \rho>0.
\EEA
For $n=m$, (\ref{we1}) leads to equality:
$\prod_{k=0}^n\sigma_{k}=\prod_{k=0}^n\spe_{k}$.

\section{Additional features of the entropy.}
\label{app_entropy}

Recall the definitions (\ref{hopo}) and (\ref{block}) of the entropy $h$ and the block
entropy $H(N)=H(\X_N,...,\X_1)$, respectively, for the stationary
process $\X$. Define:
\BEA
\label{ap3}
h(N)=H(N)-H(N-1)=H(\X_N|\X_{N-1},...,\X_1).
\EEA
$h(N)$ [sometimes called innovation entropy]
is the uncertainty of $\X_N$ given its history $\X_{N-1},...,\X_1$. It is clear that
once ${\rm lim}_{N\to\infty}\frac{H(N)}{N}$ exists, 
$h(N)$ converges to the source entropy for $N\to\infty$.
One can show that \cite{cover_thomas}
\BEA
\frac{H(N)}{N}\geq h(N)\geq h(N+1)\geq h.
\label{ap4}
\EEA
To derive the second inequality in (\ref{ap4}) note that 
the stationarity and the entropy reduction due to conditioning imply
\BEA
\label{ap5}
h(N)=H(\X_N|\X_{N-1},...,\X_1)=H(\X_{N+1}|\X_{N},...,\X_2)\geq H(\X_{N+1}|\X_{N},...,\X_1)=h(N+1).
\EEA
The first inequality in (\ref{ap4}) is shown as follows. 
\BEA
\frac{H(N)}{N}=\frac{1}{N}H(\X_1)+
\frac{1}{N}
\sum_{i=2}^N H(\X_i|\X_{i-1},...,\X_1)
\geq \frac{1}{N}\sum_{i=1}^N H(\X_N|\X_{N-1},...,\X_1)
=h(N),
\label{gomes}
\EEA
where the first equality is the obvious chain rule for the conditional
information, while the second inequality in (\ref{gomes}) follows from
the stationarity $H(\X_1)=H(\X_N)$, and then from the same reasoning as
in (\ref{ap5}).  The last inequality in (\ref{ap4}) is now obvious. 

The meaning of $\frac{H(N)}{N}\geq h\equiv {\rm
lim}_{N\to\infty}\frac{H(N)}{N}$ is that taking into account all the
correlations decreases the entropy. In a related context, $h(N)\geq
h(N-1)$ means that the innovations decrease under accumulation of
experience. This inequality can be employed for putting an upper bound
for $H(N+1)$ in terms of $H(N)$ and $H(N-1)$:
\BEA
2H(N)-H(N-1)\geq H(N+1)\geq H(N).
\EEA
Note also that $H(N+1)=H(N)+h(N+1)\leq H(N)+\frac{H(N+1)}{N+1}$
leads to 
\BEA
\frac{H(N+1)}{N+1}\leq \frac{H(N)}{N},
\EEA
i.e., the uncertainty per step decreases when increasing $N$.

\section{Ergodic features of the singular values for a random matrix product.}
\label{osel}

Let us recall some important features of the Lyapunov exponents
of the random matrix product (\ref{mubarak}). Employ the known relation between the 
singular values of $AB$ versus those of $A$ and $B$ \cite{horn}
\BEA
\label{togo}
\prod_{k=0}^m\sigma_k[AB] \leq \prod_{k=0}^m\sigma_k[A]\sigma_k[B],
\EEA
where $0\leq m\leq L-1$, and
where the ordering (\ref{grad}) is assumed: $\sigma_0[A]\geq \sigma_1[A]\geq\ldots$.

Now recall definitions (\ref{barak}, \ref{laokon}).
Applying (\ref{togo}) with $m=0$ to $\mathbb{T}(\bx)$ we get ($M<N$)
\BEA
\label{subo}
\ln\sigma_0[\mathbb{T}(\bx)]
\leq \ln\sigma_0[\mathbb{T}(\bfx_{M-1\ldots 1})]
+\ln\sigma_0[\mathbb{T}(\bfx_{N\ldots M})].
\EEA
Thus, $\ln\sigma_0[\mathbb{T}(\bx)]$ is sub-additive.  Together with
the assumptions {\it i)}, {\it ii)} and {\it iii)} of section \ref{yahud},
Eq.~(\ref{subo}) ensures
the applicability of the sub-additive ergodic theorem
\cite{kingman,steele}.  This leads (for $N\to\infty$) 
to the probability-one convergence (\ref{puk}):
\BEA
\label{moto}
-\frac{1}{N}\ln\sigma_k[\mathbb{T}(\bx)]\to \mu_k,
\EEA 
for $k=0$. Applying in the same way (\ref{togo}) with $m=1$ to $\mathbb{T}(\bx)$,
we use the sub-additivity for $\ln\left(\sigma_0[\mathbb{T}(\bx)]
\sigma_1[\mathbb{T}(\bx)]\right)$, deduce (\ref{puk}) for $k=1$, and so
on.  It is clear that we could not employ the sub-additivity 
directly for $\spe_k[\T(\bx)]$ (modules of the eigenvalues), 
since they in general do not satisfy to anything like (\ref{togo}). 

The sub-additive ergodic theorem is related to the
additive (Birkhoff-Khinchin) ergodic theorem that claims the existence
(with probability one) of a similar limit for a function
$\frac{1}{N}\sum_{k=1}^N f[\X_k]$ of the stationary random
process $\X=\{\X_1,\ldots,\X_N,\ldots\}$ \cite{steele}. 

\section{Eigenvalues and singular values 
of the random matrix product.}
\label{kovtun}

Recall section \ref{golem} and the main question posed there: when the modules of the eigenvalues of
the matrix product $\T(\bx)$ are equal, for $N\gg 1$, to the singular
values of $\T(\bx)$.

As shown by (\ref{lelak}),
for $N\gg 1$ we can keep the dependence on $N$ only in
the singular values of $\T$. (We simplified notations as $\T(\bx)=\T$.)
First assume that 
$\T$ is a $2\times 2$ matrix. Write the singular value decomposition (\ref{hop})
for $\T$ as
\BEA
\label{koyot}
\T= \left(\begin{array}{rr}
e^{-N\mu_0} & 0~~ \\
\\
0~~~~ & e^{-N\mu_1} \\
\end{array}\right)U,\qquad U=
\left(\begin{array}{rr}
a & b \\
\\
c & d \\
\end{array}\right),
\EEA
where $e^{-N\mu_0}$ and $e^{-N\mu_1}$ [with $\mu_0<\mu_1$] are the singular values of
$\T$, and where the matrix $U$ can be taken real, since $\T$ is real.
Thus $U$ is orthogonal: $ab+cd=0$, $a^2+c^2=b^2+d^2=1$, $ad-bc=\pm 1$. 

For the modules of the eigenvalues of $\T$ in (\ref{koyot}) one finds
\BEA
\spe_0=|a| e^{-N\mu_0}+\frac{|bc|}{|a|}e^{-N(\mu_1-\mu_0)}+\ldots,
\qquad
\spe_1=\frac{1}{|a|} e^{-N\mu_1}-\frac{2|bc|}{|a|^3}\,e^{-N(2\mu_1-\mu_0)}+\ldots.
\EEA
If $|a|\not =0$, the singular values of $\T$ coincide with the absolute
values of its eigenvalues for $N\gg 1$ \cite{or}: the terms ${\cal
O}(e^{-N(\mu_1-\mu_0)})$ and ${\cal O}(e^{-N(2\mu_1-\mu_0)})$ are
negligible and $\ln|a|$ is also neglected inside of the exponents as
compared to $N\mu_0$ and $N\mu_1$. 

This conclusion changes for $a=0$ (and thus $d=0$ since $U$ is
orthogonal).  Now the modules of the eigenvalues coincide with each
other and are equal to $e^{-N(\mu_1+\mu_2)/2}$ which is different from
the singular values. 

The next example is $3\times 3$ matrix $\T$ with the determinant equal to zero:
\BEA
\label{koyot1}
\T= \left(\begin{array}{rrr}
e^{-N\mu_0} & 0~~ & 0~~\\
\\
0~~~~ & e^{-N\mu_1} & 0~~\\
\\
0~~~~ & 0~~ & 0~~\\
\end{array}\right)U,\qquad U=
\left(\begin{array}{rrr}
a & b & e\\
\\
c & d & f\\
\\
x & y & z
\end{array}\right),
\EEA
where $e^{-N\mu_0}$ and $e^{-N\mu_1}$ [with $\mu_0<\mu_1$] are two
non-zero singular values of $\T$, and where the matrix $U$ is
orthogonal. Note that provided the third Lyapunov exponent $\mu_2$ is
larger than $\mu_1$ (and provided we do not use the orthogonality features
of the matrix $U$ in (\ref{koyot1})), the considered example is
sufficiently general. 

Since ${\rm det}\,\T =0$, the third singular value of $\T$ is zero.
The third eigenvalue of $\T(\bx)$ is also equal to zero,
while for the absolute values of the remaining eigenvalues we have from
(\ref{koyot1})
\BEA
\spe_0=|a| e^{-N\mu_0}+{\cal O}(\frac{1}{|a|}\,e^{-N(\mu_1-\mu_0)}),
\qquad
\spe_1=\frac{|ad-bc|}{|a|} e^{-N\mu_1}+{\cal O}(e^{-N(2\mu_1-\mu_0)}).
\EEA
If $|ad-bc|\not= 0$, the singular values $e^{-N\mu_0}$ and $e^{-N\mu_1}$
coincide [for $N\gg 1$] with the modules of the eigenvalues.  For
$|ad-bc|= 0$ the second eigenvalue of $\T$ is equal to zero, while
the second singular value is non-zero. However, the first Lyapunov exponent is
still equal to the spectral radius (module of the first eigenvalue) if $a\not=0$. The
latter two quantities are not equal for $a=0$. Now the modules of both
eigenvalues of $\T(\bx)$ reduce to $\sqrt{|bc|}\,e^{-N(\mu_1+\mu_2)/2}$. 

Using the examples (\ref{koyot}, \ref{koyot1}) we got a sufficient
condition for deciding whether the maximal singular value of $\T$
is equal to the module of the corresponding eigenvalue. It is that the
absolute values of the two leading eigenvalues of $\T$ are different.

\section{Zeta-function and periodic orbit expansion. }
\label{zeta_app}

\subsection{Structure of periodic orbits.}

Define formally
\BEA
\label{kalibr_z}
Z_m=\sum_{i_1,...,i_m=1}^M \phi [A_{i_1}...A_{i_m}],
\EEA
where $A_1,...,A_M$ are matrices, and where $\phi[.]$ is a function that turn its matrix argument
to a number. We assume that the following features hold for $\phi$
($d$ is a positive integer):
\BEA
\label{kalibr_phi}
\phi[A^d]=\phi^d[A], \qquad \phi[AB]=\phi[BA].
\EEA
Using these features one can prove for $Z_m$ the following formula \cite{ruelle}:
\BEA
\label{kaban}
Z_m=\sum_{n|m}\,\,\,\sum_{(\gamma_1,...,\gamma_n) \in {\rm Per}(n) } n\left[\,
\phi[A_{\gamma_1}...A_{\gamma_n}]\,
\right]^{\frac{m}{n}},
\EEA
where $\sum_{n|m}$ means that the summation goes over all $n$ that
divide $m$, e.g., $n=1,2,4$ for $m=4$.  Here ${\rm Per}(n)$ contains
sequences 
\BEA
\Gamma=(\gamma_1,...,\gamma_n) 
\EEA
selected according to the following rules: {\it i)} $\Gamma$ turns to
itself after $n$ successive cyclic permutations, but does not turn to
itself after any smaller (than $n$) number of successive cyclic
permutations; {\it ii)} if $\Gamma$ is in ${\rm Per}(n)$, then ${\rm
Per}(n)$ contains none of those $n-1$ sequences obtained from $\Gamma$
under $n-1$ successive cyclic permutations. 

Assume that $M=2$, which means that the matrices $A_i$ can take two
values $A_1=1$ and $A_2=2$.  With examples of ${\rm Per}(n)$ given in
Table~\ref{tab1}, the proof of (\ref{kaban}) is straightforward. 

\begin{table}
\caption{The elements of ${\rm Per}(p)$ for $p=1,...,5$ and $M=2$.
As compared to (\ref{barak}) we denoted $T(x_1)=1$ and $T(x_2)=2$.
It is seen that ${\rm Per}(1)$
contains two elements, since the cyclic permutation is trivial. ${\rm
Per}(2)$ contains a single element $12$, since $11$ and $22$ remain
invariant under a single cyclic permutation, while $BA$ is obtained from
$AB$ via a single cyclic permutation. Besides the obvious sequences
$1111$ and $2222$, ${\rm Per}(4)$ does not include the sequences $1212$ and
$2121$ which stay invariant after two successive cyclic permutations. In
${\rm Per}(5)$ we first meet different elements that have the same
overall number of $1$'s and $2$'s, e.g., $12121$ and $11122$. 
}
\begin{tabular}{|c||c|}
\hline
$p$             & ${\rm Per}(p)$                   \\
\hline
\,\, $1$   \,\, & $1$, $2$                    \\
\hline
\,\, $2$   \,\, & $12$  \\
\hline
\,\, $3$   \,\, & $122$, $211$  \\
\hline
\,\, $4$   \,\, & $1222$, $2111$, $1122$  \\
\hline
\,\, $5$   \,\, & $12222$, $21111$, $11222$,\\ 
                &  $22111$, $12121$, $21212$  \\
\hline
\,\, $6$   \,\, & $122222$, $112222$, $111222$,\\ 
                &  $111122$, $111112$,          \\
                &  $112212$, $221121$          \\
                &  $111212$, $222121$.          \\
\hline
\end{tabular}
\label{tab1}
\end{table}

\begin{table}
\caption{The elements of ${\rm Per}(p)$ for $p=1,...,4$ and $M=3$.
}
\begin{tabular}{|c||c|}
\hline
$p$             & ${\rm Per}(p)$                   \\
\hline
\,\, $1$   \,\, & $1,2,3$                    \\
\hline
\,\, $2$   \,\, & $12$, $13$, $23$  \\
\hline
\,\, $3$   \,\, & $122$, $211$, $233$  \\
\,\,       \,\, & $322$, $133$, $311$  \\
\,\,       \,\, & $123$, $132$         \\
\hline
\,\, $4$   \,\, & $1222$, $2111$, $1122$,\\ 
                & $2333$, $3222$, $2233$  \\
\,\,       \,\, & $1333$, $3111$, $1133$  \\
\,\,       \,\, & $1123$, $1132$, $1213$  \\
\,\,       \,\, & $2213$, $2231$, $2321$  \\
\,\,       \,\, & $3312$, $3321$, $3231$  \\
\hline
\end{tabular}
\label{tab2}
\end{table}

\subsection{The inverse zeta-function and derivation of Eq.~(\ref{castro}).}
\label{fidelito}

The inverse zeta function is defined as $\xi(z)=\exp\left[-\sum_{m=1}^\infty
\frac{z^m}{m}Z_m\right]$, where $Z_m$ is given by (\ref{kalibr_z}).
Employing (\ref{kaban}) and introducing notations $p=n$,
$q=\frac{m}{n}$, we transform $\xi(z)$ as
\BEA
\xi(z)=\exp\left[\,-
\sum_{p=1}^\infty \,
\sum_{ \Gamma \in {\rm Per}(p) }\, \sum_{q=1}^\infty \frac{z^{pq}}{q}
\left(
\phi[A_{\gamma_1}...A_{\gamma_p}]\,
\right)^q
\right].
\label{trevoga_1}
\EEA
the summation over $q$ in (\ref{trevoga_1}) is taken as
\BEA
\sum_{q=1}^\infty \frac{z^{pq}}{q}
\left(
\phi[A_{\gamma_1}...A_{\gamma_p}]\,
\right)^q=-\ln\left[
1-z^p \phi[A_{\gamma_1}...A_{\gamma_p}]\,
\right].
\EEA

We shall then finally get \cite{artuso,ruelle}:
\BEA
\xi(z)=\prod_{p=1}^\infty \,
\prod_{\Gamma \in {\rm Per}(p) }\, 
\left[
1-z^p \phi[A_{\gamma_1}...A_{\gamma_p}]\,
\right].
\EEA

\subsection{How to generate the elements of ${\rm Per}(p)$ via Mathematica 5.}
\label{mato}

The elements of ${\rm Per}(p)$ presented in Tables \ref{tab1} and
\ref{tab2} were generated by hands. For larger $p$ it is more convenient
to generate these elements via Mathematica 5.  Below we assume that the
reader knows Mathematica at some average level.  First one should run
the package of combinatoric functions:
\BEA
\texttt{<<DiscreteMath`Combinatorica` 
}
\label{mangust}
\EEA
Next one defines the function $\texttt{ListNecklaces2[c\_List, n\_Integer?Positive]}$ \cite{list}, the
first argument of which is a list, e.g., $\texttt{ \{ A,B \} }$, while the second argument is
a positive integer.
\BEA
&&\texttt{AllCombinations[x\_List, n\_Integer?NonNegative] } \nonumber\\
&&\texttt{
:= 
  Flatten[Outer[List, Sequence \@\@ Table[x, \{n\}]], n - 1]; }\nonumber\\
&&\texttt{ListNecklaces2[c\_List, n\_Integer?Positive] := 
  Module[\{\}, }\nonumber\\
&&\texttt{
    Return[OrbitRepresentatives[CyclicGroup[n], AllCombinations[c, n]]]];
}
\label{neru}
\EEA
The definition of $\texttt{ListNecklaces2}$ proceeds via an auxiliary function
$\texttt{AllCombinations}$. All other functions in (\ref{neru}) are contained in the
package (\ref{mangust}).

Upon running $\texttt{ListNecklaces2[c, p]}$ one gets the elements of
${\rm Per(p)}$ together with those sequences $(\gamma_1,...,\gamma_p)$
that remain invariant under $\bar{p}$ successive cyclic permutation,
where $p/\bar{p}$ is an integer.  For our purposes we meed only the
sequences which are invariant with respect to $p$ cyclic permutation,
and are not variant with respect to cyclic permutations with any smaller
$\bar{p}$.  So our next task is to get rid of those parasitic sequences,
which stay invariant with respect to $\bar{p}$ cyclic permutations with
$\bar{p}<p$. To this end we designed a straightforward Mathematica
program that by the direct enumeration detects and eliminates the
parasitic sequences [obviously, nothing special has to be done for
simple numbers like $p=3,5,7,11,13$]. The drawback of this program is
that for each $p$ in ${\rm Per}(p)$ one has to adjust the details of
this program. Anyhow, we were not able to enforce Mathematica 5 to
generate the elements of ${\rm Per}(p)$ directly. 

Here is an example of the above scheme: $\texttt{ListNecklaces2[\{A,B\}, 3]}$ generates
a list of lists:
\BEA
\texttt{\{ \{ A,A,A \}, \{ A,A,B\}, \{ A,B,B\}, \{ B,B,B\} \}}.
\EEA
After elimination of the parasitic sequences this results in
\BEA
\texttt{Y}=\texttt{\{ \{ A,A,B\}, \{ A,B,B\}  \}},
\EEA
where we introduced a shorthand $\texttt{Y}$. Now employing the construction
\BEA
\texttt{
Apply[Times, Map[  f[\#] \&, Apply[Dot, Y, 1] ] ]
},
\label{camaradas}
\EEA
where $\texttt{f}$ is an arbitrary function,
one gets
\BEA
\texttt{f[A.A.B] f[A.B.B]}.
\EEA
The construction (\ref{camaradas}) is useful when recovering the formulas for $\phi_k$ for large
values of $p$.


\begin{thebibliography}{99}

\bibitem{rabiner_review}
L. R. Rabiner,  Proc. IEEE, {\bf 77}, 257-286, (1989).

%A tutorial on hidden Markov models and selected
%applications in speech recognition

\bibitem{ephraim_review}
Y. Ephraim and N. Merhav, IEEE Trans. Inf. Th., {\bf 48}, 1518-1569, (2002).

%%Hidden Markov processes

\bibitem{signal}
M. Crouse, R. Nowak and R. Baraniuk, IEEE Tran. Signal
Process., {\bf 46}, 886 (1998).

%% Wavelet-based statistical signal processing using hidden Markov models


\bibitem{dna} T. Koski, {\it Hidden Markov Models for Bioinformatics} 
(Kluwer, Academic Publishers, Dordrecht, 2001).

P. Baldi and S. Brunak, {\it Bioinformatics} (MIT Press, Cambridge, USA, 2001).

%%G. A. Churchill, Bull. Math. Biology, {\bf 51}, 79 (1989).

%% Stochastic models for heterogeneous DNA sequences

%%A. V. Lukashin and M. Borodovsky, 
%%Nucleic Acids Res, {\bf 26}, 1107 (1998).

%% New solutions for gene finding

\bibitem{ash}R. Ash, {\it Information Theory} (Interscience
Publishers, NY, 1965).

\bibitem{cover_thomas}
T. M. Cover and J. A. Thomas, {\it Elements of Information Theory},
(Wiley, New York, 1991).

\bibitem{blackwell}
D. Blackwell, {\it The entropy of functions of finite-state
markov chains}, in Trans. First Prague Conf. Inf. Th., Statistical
Decision Functions, Random Processes, p. 13 
(Pub. House Chechoslovak Acad. Sci., Prague, Czechoslovakia, 1957).

\bibitem{strat}R.L. Stratonovich, {\it Information Theory} (Sovietskoe Radio, Moscow, 1976) (In Russian).

\bibitem{reza}M. Rezaeian, {\it Hidden Markov Process: A New Representation, Entropy Rate
and Estimation Entropy}, arXiv:cs.IT/0606114.

\bibitem{birch} I.J. Birch, Ann. Math. Stat. {\bf 33}, 930 (1962).

\bibitem{jacquet}
P. Jacquet, G. Seroussi, and W. Szpankowski, {\it On the Entropy
of a Hidden Markov Process}, Int. Symp. Inf. Th. p. 10,
Chicago, IL, 2004.

\bibitem{gold} T. Holliday, A. Goldsmith and P. Glynn,
IEEE Trans. Inf. Th. {\bf 52}, 3509 (2006).

\bibitem{weissman}
E. Ordentlich and T. Weissman, IEEE Trans. Inf. Th., {\bf 52}, 19 (2006).
%% On the optimality of symbol by symbol filtering and denoising

\bibitem{egner}
S. Egner {\it et al.}, {\it On the Entropy
Rate of a Hidden Markov Model}, Int. Symp. Inf. Th., p. 12,
Chicago, IL (2004).

\bibitem{zuk_jsp}
O. Zuk, I. Kanter and E. Domany,  J. Stat. Phys. {\bf 121}, 343 (2005).
%% The Entropy of a Binary Hidden Markov Process

\bibitem{zuk_aizenman}
O. Zuk, I. Kanter, E. Domany and M. Aizenman,  IEEE Signal Processing Letters, {\bf 13}, 517 (2006).

\bibitem{chigan}P.Chigansky, {\it 
The Entropy Rate of a Binary Channel with Slolwly Varying Input}, arXiv:cs/0602074.



%%\bibitem{mushkin} M. Mushkin and I. Bar-David,  IEEE Trans. Inf. Theory, {\bf IT-35}, 1277 (1989).
%% Capacity and Coding for the Gilbert-Elliot Channel.


\bibitem{horn}R. A. Horn and C. R. Johnson, {\it Matrix Analysis} (Cambridge University Press, New Jersey,
USA, 1985).



%%\bibitem{triple}W. J. McGill, IEEE Trans. Inf. Theory, {\bf 4}, 93 (1954).

%%T. S. Han, Information and Control, {\bf 46}, 26 (1980).

%%T. Tsujishita, Advances in Applied Mathematics, {\bf 16}, 269 (1995).

%%A. Jakulin and I. Bratko, arXiv:cs.AI/0308002.

%%\bibitem{triple_f}H. Matsuda, Phys. Rev. E, {\bf 62}, 3096 (2000).


%%\bibitem{neural} N. Brenner {\it et al.}, Neural Comput., {\bf 12} 1531 (2000).
%% Synergy in a neural code. 

%%E. Schneidman {\it et al.}, Journal of Neuroscience, {\bf 23}, 11539 (2003).

%% Synergy, redundancy, and independence in population codes.

\bibitem{kingman}J.F.C. Kingman, Ann. Probab. {\bf 1}, 883 (1973).

\bibitem{steele} J.M. Steele, Annales de l'I.H.P. B, {\bf 25}, 93 (1989).

\bibitem{crisanti} A. Crisanti, G. Paladin and A. Vulpiani, 
{\it Products of Random Matrices in Statistical Physics}, Springer Series in
Solid State Sciences, Vol. 104, (Springer, Berlin, 1993).

\bibitem{goldsheid}L.Y. Goldsheid and G.A. Margulis, Russ. Math. Surveys {\bf 44}, 11 (1989).

%%\bibitem{furst}H. Furstenberg and H. Kesten, Ann. Math. Statist. {\bf 31}, 457 (1960).

\bibitem{or}S.A. Orszag, P.L. Sulem and I. Goldirsch, Physica D {\bf 27}, 311 (1987).

\bibitem{kontorovich}L. Kontorovich, {\it Measure Concentration of Hidden Markov Processes},
arXiv:math/0608064.

%%\bibitem{lindgren}K. Eriksson and K. Lindgren, Physica Scripta {\bf 35}, 388 (1987).
%%K. Lindgren, Phys. Rev. A {\bf 38}, 4794 (1988).

%%\bibitem{schreiber}T. Schreiber, Phys. Rev. Lett. {\bf 85}, 461 (2000).

%%\bibitem{preprint}R. Renner and U. Maurer, preprint. 

\bibitem{artuso} R. Artuso. E. Aurell and P. Cvitanovic, Nonlinearity {\bf 3}, 325 (1990).

P. Cvitanovic, Phys. Rev. Lett. {\bf 61}, 2729 (1988).

\bibitem{ruelle}
D. Ruelle, {\it Statistical Mechanics, Thermodynamic Formalism}, (Reading, MA: Addison-Wesley, 1978).

%%%M. Pollicott, Inv. Math. {\bf 85} (1986).



\bibitem{mainieri}R. Mainieri, Chaos {\bf 2}, 91 (1992).

\bibitem{aurell}E. Aurell, J. Stat. Phys., {\bf 58}, 967 (1990).


\bibitem{gaspar} 
J. Nielsen, {\it Lyapunov exponents for products of random matrices}, preprint available at
http://citeseer.ist.psu.edu/438423.html.


%% \bibitem{oseledec}V. I. Oseledec, Trans. Moscow Math. Soc., 197 (1968).
%%  A multiplicative ergodic theorem. Lyapunov characteristic numbers for
%%  dynamical systems.

\bibitem{arnold}
L. Arnold, V. M. Gundlach and L. Demetrius, Ann. Appl. Probab., {\bf 4}, 859 (1994).

%%Evolutionary formalism for products of positive random matrices


\bibitem{peres}
Y. Peres, Ann. Inst. H. Poincare Probab. Statist., {\bf 28}, 131 (1992).

%%Domains of analytic continuation for the top Lyapunov exponent

\bibitem{han_markus}G. Han and B. Markus, IEEE Trans. Inf. Th., {\bf 52}, 5251, (2006).

\bibitem{list} I learned about the function $\texttt{ListNecklaces2}$ from the e-mail
exchange presented in 
http://forums.wolfram.com/student-support/topics/6401

\bibitem{lolo} L. Gurvits and J Ledoux,
Linear Algebra and Applications, {\bf 404}, 85 (2005).

%%Markov property for a function of a Markov chain: A linear algebra approach 

\bibitem{karl}K. Petersen, {\it Lectures on Ergodic Theory}, available from
http://www.math.unc.edu/Faculty/petersen/lecturespdf.pdf

%%Y. Sinai, {\it Topics in ergodic theory} (Princeton University Press, Princeton, 1994).

\end{thebibliography}
\end{document}